\documentclass[prb,showpacs,preprintnumbers,amsmath,aps]{revtex4}

\newcommand{\olcite}[1]{[\onlinecite{#1}]}
\usepackage{graphicx}
\usepackage{amssymb}
\usepackage{color}
\usepackage{subfig}
\begin{document}

\title[Random Pinning Glass transition]{Random Pinning Glass Transition: Hallmarks, Mean-Field Theory and Renormalization Group Analysis}

\author{Chiara Cammarota}
\address{Institut de Physique Th\'eorique
Orme des Merisiers batiment 774
Point courrier 136
CEA/DSM/IPhT, CEA/Saclay
F-91191 Gif-sur-Yvette Cedex}
\address{LPTMC, Tour 12-13/13-23, Bo"te 121, 4, Place Jussieu, 75252 Paris Cedex 05, France}

\author{Giulio Biroli}
\address{Institut de Physique Th\'eorique
Orme des Merisiers batiment 774
Point courrier 136
CEA/DSM/IPhT, CEA/Saclay
F-91191 Gif-sur-Yvette Cedex}

\begin{abstract}
We present a detailed analysis of glass transitions induced by pinning particles at random 
from an equilibrium configuration. We first develop a mean-field analysis based on the study
of p-spin spherical disordered models and then obtain the three dimensional critical behavior 
by the Migdal-Kadanoff real space renormalization group method. We unveil the important physical 
differences with the case in which particles are pinned from a random (or very high temperature)
configuration.  We contrast the pinning particles approach to the ones based on biasing 
dynamical trajectories with respect to their activity and on coupling to equilibrium configurations. Finally, we 
discuss numerical and experimental tests.  
\end{abstract}

\maketitle

\section{Introduction}

A universal feature of super-cooled liquids is the huge growth of relaxation time when 
temperature is decreased below the melting point.
In a restrict temperature window the relaxation time-scale increases by many orders of magnitude.
Starting from the microscopic dynamic time-scale of simple liquids (picoseconds), the relaxation time increases so steeply 
that below a certain temperature, $T_g$, it  exceeds the observation time scale (hours).
Eventually, for $T<T_g$, it is not possible to equilibrate the system anymore: the sample is frozen in an amorphous solid called glass~\cite{Cav_rev, BeBi_rev}.\\
Despite long lasting theoretical and experimental efforts, a definitive explanation for the origin 
of the dynamical slowing down in supercooled liquids is still lacking.
The main theoretical question is what kind of critical behavior---if any---is responsible for the 
slow dynamics. Is it of thermodynamic nature, e.g. due to a Random First Order Transition (RFOT) at a temperature $T_K<T_g$~\cite{KTW,LubWol_rev} 
or to an avoided defect-mediated transition~\cite{Gilles}, or is it instead purely dynamical~\cite{GotSjo_rev,KCMGlass,ChaGar_rev}?
In the case of usual second-order phase transitions, 
critical properties emerge and can be experimentally resolved within the critical region only, {\it i.e.} when time and length scales become much larger than the their microscopic counterparts. 
Unfortunately, in the case of the glass transition, all evidences point toward a modest increase of the length-scales associated to collective behavior, even close to $T_g$~\cite{BeBi_rev}. This is likely due to an exponential relation between time and length-scales different from the usual power law one found in standard critical phenomena~\cite{BeBi_rev}. Indeed, approaching $T_g$ 
the relaxation time increases by several orders of magnitude, but typical length-scales increase by no more than a few inter-particle distances (ten at most). The consequences of this state of affairs is that: (1) it is not possible to observe genuine critical behavior, (2) testing and comparing different dynamic and static theories is very difficult, (3) it is not known whether a true phase transition takes place at a finite temperature $T_K$ below $T_g$ or, instead, 
the transition is avoided or present at $T=0$ only.\\
We recently proposed a way to partially solve these problems and to find 
and characterize the ideal glass transition at $T_K$, if present~\cite{CB_RPGT}.
We proposed that ideal glass transitions can be induced in supercooled liquids at temperature higher than 
$T_g$ by a suitable random perturbation:
The idea is to block a fraction $c$ of particles in the positions they have in a chosen (but arbitrary) equilibrium 
configuration at temperature $T$ and study the thermodynamics of the remaining free particles.
Using the RFOT approach we found that in a finite range of temperatures, $T\in[T_K,T_h]$, a true ideal glass transition, dubbed Random Pinning Glass Transition (RPGT), is expected to occur at finite concentration $c_K(T)$ of pinned particles. 
This transition for $c\uparrow c_K(T)$ has the same features as the ideal glass transition for $T\downarrow T_K$.
However, it can be studied much more thoroughly.
The major advantage is that, as we shall explain later, the ideal glass phase is known in this case and, hence, it is possible to equilibrate the system, not only below $c_K(T)$ in the liquid phase, but also above $c_K(T)$ in the amorphous glass phase. For example, a standard local Monte Carlo or molecular dynamics can correctly probe the equilibrium measure on time scales not diverging with the system size in the glass phase. 
Therefore, for the first time, the ideal glass transition can be approached both from the liquid and from the glass phase. Showing the existence and the critical properties of the transition is not left to doubtful 
extrapolations but can be ascertained by using the finite size scaling machinery developed for standard phase-transitions. Moreover, information on the presence, the position in temperature and the critical properties of the glass transition of the unperturbed system might also be numerically or even experimentally grasped 
by extrapolating at $c=0$ the analysis performed at finite $c$ for $T>T_g$. \\
The main aspects of the physical mechanism behind random pinning glass transitions and some results 
have been announced in~\olcite{CB_RPGT}.
In this paper we present a comprehensive derivation and analysis of the theoretical predictions, including in particular the phase diagram in the $c-T$ plane and the critical properties. The latter have been obtained through a non-perturbative renormalization group method that we describe in detail. In this work we also compare our approach to alternative ones developed by Franz and Parisi~\cite{FraPar97} and Chandler and Garrahan and co-workers~\cite{ChaGar_rev,sensemble,vanwijland}.  Despite similarities, especially concerning the phase diagrams, these
two approaches turn out to be quite different: in particular they predict a first-order phase transition line and not a {\it bona-fide} glass transition line, as in our case. The difference between pinning
from an equilibrated and from a random configuration is also addressed in this work. We shall show that the latter procedure leads to quite different 
phase diagram and physical behaviors.  

The procedure of pinning particles from an equilibrium configuration plays a key role in our work. It appeared in the
literature already a decade ago in \olcite{SKBP} and \olcite{Kim1}. After the suggestion in~\olcite{BouBir}, it became an important tool for testing theoretical predictions related to the ideal glass transition and probe medium range amorphous order~\cite{Cavagnacavity1,BBCGV}.
A number of different geometries for the set of pinned particles have been studied~\cite{BBCGV,SauTar10, ChChTa12,BeKo_PS,HocRei12,GrTrCa12}: 
the cavity, the two-wall and the one-wall geometry and pinning a fraction of particles chosen 
at random. 
Although already in~\olcite{SKBP,Kim1} it was observed that this kind of procedures induces a significant slowing 
down of the dynamics of the remaining free particles, only recently it was claimed by numerical simulations that the impressive increase in relaxation time may be related to a true 
divergence~\cite{BeKo_PS, ProKar11}. 
In our previous work~\olcite{CB_RPGT} we show that this divergence is theoretically expected within RFOT theory and it is associated with an ideal glass transition, analogous to the one at $T_K$, occurring 
even at rather high temperatures, in some cases even close to the onset of glassy dynamics.
The predictions of approaches other than RFOT theory concerning the effect of pinning particles appear to be quite different ranging 
from a gradual cross-over (instead of a singularity), as it is the case for some dynamic facilitation models  \cite{JacBer12}, to no characteristic effect at all. 
For this and other reasons,
 the pinning particles procedure is a natural framework  where different theories of the glass transition can be fruitfully challenged. 

The structure of the manuscript is the following: in Sec. 2 we review the phenomenological approach  proposed in Ref.~\onlinecite{CB_RPGT} to explain random pinning glass transitions. The bulk of the paper 
that contains the mean-field analysis, the RG study and a toy model corresponds to Secs. 3, 4 and 5 respectively. 
Sec. 6 is devoted to a comparison between the random pinning approach and Franz-Parisi and Chandler-Garrahan ones. Finally, in Sec. 7 we present a general discussion, highlight the main advantages of the pinning particles procedure to study the glass transition and we suggest numerical and experimental tests. 

\section{The random pinning glass transition: a phenomenological approach}
\label{phenomenology}
In this section, we recall the phenomenological arguments we presented in~\olcite{CB_RPGT} to explain 
random pinning glass transitions. 
Since our basic framework is RFOT theory we have first to present its main
physical ingredient, which is the competition between two thermodynamic quantities: 
the configurational entropy $s_c(T)$, which is the entropy per unit volume associated to the multiplicity of the amorphous phases in which
a super-cooled liquid can freeze, and the interface free-energy cost $\Delta F$ of surfaces separating regions arranged in different amorphous phases.
On length-scale $\ell$ these two quantities respectively scale as $s_c(T)\ell^d$ and $\Upsilon \ell^{\theta}$, where 
$\theta$ is equal or less than $d-1$ ($d$ is the spatial dimension), 
$\Upsilon$ is the so-called amorphous surface tension. 
By lowering the temperature $s_c(T)$ decreases; it is assumed to vanish with non-zero first derivative at a finite temperature $T_K$\footnote{Actually, this assumption is not necessary
to explain the physical behavior of super-cooled liquids; the configurational entropy could eventually not vanish, even though all the experimentally accessible physical behavior may be explained {\it as if} it were vanishing at a finite temperature $T_K$.}. 
This is at least what one finds in mean-field models~\cite{Cav_rev}, approximate computations~\cite{MePa_rev} and in extrapolations from numerical data~\cite{Sc05_rev,Heu08_rev}. 
By comparing the scaling with $\ell$, one finds that on small length-scales the interface free-energy cost prevents the system from freely rearranging in different amorphous phases; the entropic gain due to the rearrangement instead becomes dominant on large length-scales.
By balancing the two terms one finds a characteristic length $\ell_{PS}=(\Upsilon/Ts_c)^{1/(d-\theta)}$, called point-to-set. For $\ell<\ell_{PS}$, it is preferable for the system to be localized in a single given amorphous phase, characterized by a specific hidden amorphous order.
For $\ell>\ell_{PS}$, the amorphous order is disrupted and the system is organized in a mosaic state, a kind of micro-phase separated state in which the number of competing phases is actually huge. The relaxation time is governed, in this picture, by cooperative activated rearrangements on the length-scale $\ell_{PS}$. These involve barriers that scale with $\ell_{PS}$ and, hence, lead to a relaxation time $\tau\simeq\tau_0\exp(A \ell_{PS}^{\psi}/T)$.
Thanks to the power law divergence of $\ell_{PS}$ at $T_K$, the RFOT theory obtains a super-Arrhenius law for the relaxation time-scale of the system and, for certain combination of the exponents $\theta$ and $\psi$, the Adam-Gibbs law. Note that there is not yet a firm microscopic derivation of the values of the exponents $\theta, \psi$. Several recent works have been devoted to this issue~\cite{CCGGVexp,KaDaSa09,FleSza10} and heuristic arguments put forward in 
\olcite{KTW} suggest $\theta=\psi=\frac d 2$. See the recent book \onlinecite{bookRFOT} for more details on RFOT.\\
Let's see how does the picture change when a fraction of particles is blocked from an equilibrium configuration.
The exact procedure is the following: take an equilibrium configuration, pick at random a fraction $c$ of particles,
pin them, {\it i.e.} do not allow them to move anymore and then study the (static and dynamic) physical behavior
of the remaining unconstrained particles. Of course both their configurational entropy density, $s_c(c,T)$, and their amorphous surface tension, $\Upsilon(c,Y)$, are affected by pinning. However, we expect that the dominant change, at least close to $T_K$, is in $s_c(c,T)$ since the configurational entropy is small close to $T_K$ and the 
extra constraints due to the blocked particles, diminish it even further. 
Indeed, any blocked particle affects the possible configurations of its neighbors leading to a microscopic decrease of the effective configurational entropy density.
For small $c$ let us approximately write the new configurational entropy density as $s_c(c,T)\simeq s_c(T)-cY(T)$.
Increasing $c$  the configurational entropy monotonically decreases. For temperatures close to $T_K$, $s_c(c,T)$ is expected to vanish linearly as $s_c(c,t)\simeq Y(T)(c_K(T)-c)$ at the critical fraction $c_K(T)\simeq s_c(T)/Y(T)$.\\
By repeating the usual RFOT theory arguments but now with a configurational entropy that vanishes increasing $c$
(instead of decreasing $T$), one obtains the physical picture of an entropy vanishing transition
at $c_K(T)$, analogous to the one at $T_K$.
In the pinned system, the effective competing terms are $\Upsilon(c,T) \ell^{\theta}$ and $s_c(c,T)\ell^d$ with $\Upsilon(c,T)\simeq\Upsilon(T)$ and $s_c(c,T)<s_c(T)$. As a consequence, the liquid phase in presence of pinned particles is characterized by much larger cooperative rearranging regions of size $\ell_{PS}(c,T)=(\Upsilon(c,T)/Ts_c(c,T))^{1/(d-\theta)}\gg \ell_{PS}(T)$. The mosaic length-scale of the randomly pinned liquid
 grows when $c$ increases and eventually diverges at $c_K(T)$. Because of the relation discussed before between
$\ell_{PS}$ and the relaxation time, one expects approaching to the entropy vanishing transition a generalized
Vogel-Fulcher divergence $\tau\simeq\tau_0\exp(\ell_{PS}(c,T)^{\psi}/T)=\tau_0\exp(A(T)/(c_K(T)-c)^{\psi/(d-\theta)})$.
At higher temperature this picture breaks down for the effective amorphous surface tension term $\Upsilon(c,T)$ changes. The reason is that fluctuations are so strong at high temperature that metastable states are not defined 
anymore, {\it i.e.} the interface-free energy cost $\Upsilon(c,T)$ is expected to vanish in the high temperature regime. 
Hence, random pinning glass transitions should only exist in a finite temperature range $T_K<T<T_h$ where $T_h$ is defined as the temperature such that $\lim_{T\rightarrow T_h^-}\Upsilon(c_K(T),T)=0$.\\
One of the main conclusion one can draw from the previous arguments is that by 
approaching random pinning glass transitions by increasing $c$ toward $c_K(T)$
one should find the same physical behavior observed (or expected) for super-cooled liquids increasing $T$ towards $T_K$.
Instead, beyond the transition, there are important differences. In particular, whereas the ideal glass phase 
below $T_K$ is completely out of reach and can be only found for small systems by solving a daunting optimization problem the one beyond $c_K$ is easy to obtain: it is given by the initial configuration used to pin particles (plus vibrations around it). The physical picture is that below $c_K$ the ideal glass is just one out of the very many states sampled by the liquid; at $c_K$ when the configurational entropy vanishes it is the only one to remain available. 
All other amorphous states have a higher free energy for $c>c_K(T)$. (This picture is 
quite similar to the one worked out for "quiet planting" in random constraint satisfaction optimization problems \cite{KrZdPlanting}). 
A particularly instructive example where this mechanism can be found at work is provided by the (pinned) Random Energy Model, see \olcite{CB_RPGT} (and also \olcite{FrPaTe08}). \\
Note that the existence of a preferred state related to the initial configuration makes clear that pinning from an equilibrium configuration is very different from blocking the particle positions completely at random. In the latter case, there is no preferred configuration. In consequence, pinning acts like an external uncorrelated quenched disorder and finding
the ideal glass state above $c_K$ is expected to be as difficult as finding it below $T_K$. We shall discuss in detail this point and the comparison between the two procedures later. 

\section{Mean field theory of Random Pinning Glass Transitions}
\label{pspin}
By now it is well established that the study of the disordered spherical $p$-spin completely connected model with $p\ge3$ provides a sketchy mean-field theory for the physics of supercooled liquids~\cite{Cav_rev,BeBi_rev,BoBi_rev,KTpspin,CriSom,BoCuKuMe}
and RFOT.
The mean field replica solution of this particular spin model displays an entropy vanishing transition at $T_K$, where the system undergoes a one-step Replica Symmetry Breaking ($1$-RSB) transition~\cite{CriSom}.
In this section we use this model to obtain a mean-field theory of RPGTs. Instead of pinning particles we block
a fraction $c$ of spins from an equilibrium configuration at temperature $T$. For comparison, we also 
work out the same analysis when the spins are blocked from a high temperature (essentially random) configuration, which mimics the case of particles blocked in completely random positions.\\  
In the context of combinatorial optimization, a similar analysis has been performed in Ref.\onlinecite{SeRT09}. 
\subsection{The spherical $p$-spin completely connected model}
Let us first recall definitions and notations for the usual spherical $p$-spin model ({\it i.e.} without pinning).
Its Hamiltonian reads:
\begin{equation}
H=-\sum_{(i,j,\dots k)}J_{ij\dots k}s_is_j\dots s_k
\label{Hamilton}
\end{equation}
where the sum is performed over all the possible groups of $p$ spins in a system of $N$ spins, $J_{ij\dots k}$ are i.i.d. Gaussian random variables with zero mean and variance $\sigma_J=p!/2N^{p-1}$, and the spin $s_i$ are continuous variables bounded by the following spherical constraint: $N=\sum_{i=1}^Ns_i^2$.
The mean intensive free-energy, $\mathcal{F}(T)$, averaged over the possible choices of the disorder, is obtained introducing $n$ replicas of the system. Performing standard replica manipulations~\cite{Cavspin_rev} we find (in the following we rescale $T$ in units of $k_B$ and use the notation $\overline{\phantom{a}.\phantom{a}}$ for the average over the quenched disorder):
\begin{equation}
\mathcal{F}(T)=-\frac{T}{N}\overline{\log Z}=-\frac{T}{N}\lim_{n\rightarrow0}\frac{\log \overline{Z^n}}{n}=-\lim_{n\rightarrow0}\lim_{N\rightarrow\infty} \frac{T}{Nn}\log \int dQ_{a,b}\exp[-NS(Q_{ab};T)]
\label{free-energy}
\end{equation}
where $Q_{ab}$ is a $n\times n$ symmetric matrix and
$$S(Q_{ab})=-\frac{1}{4T^2}\sum_{a,b}^nQ_{ab}^p-\frac{1}{2}\log\det(Q_{ab})\ .$$
The saddle point solution for $Q_{ab}$ in (\ref{free-energy}) is a matrix with $n/m$ non-zero diagonal blocks of $m$ x $(m-1)$ elements equal to $q_1$, and $m$ diagonal unitary elements.
The saddle point equations determine the equilibrium value of $q_1$ and $m$. 
At a temperature $T_K$ the free-energy has a singularity below which the equilibrium value of $q_1$ jumps from $0$ to a non-zero value $q_1^{EA}$ and $m$ starts to decrease from $1$, this is the signature of the $1$RSB transition.
The other characteristic temperature in the model is the dynamic transition temperature $T_d$. It marks the presence of a dynamical transition akin to the one found in Mode Coupling Theory; $T_d$ can be obtained as the highest temperature where one finds a stationary point of the action characterized by a non-zero value of $q_1$ for $m=1$~\cite{CriSom,Cavspin_rev}.\\
In the next two subsections we extend the usual replica computations in the case where a finite fraction $c$ of spins are blocked according to the two different protocols cited above.

\subsection{Random pinning from equilibrium configurations}
\label{MF_EQ}

The free-energy of a $p$-spin model with a finite fraction of spins blocked from an equilibrium configuration can be obtained through a double average process. For each choice of the quenched disorder, we have to average over the possible different equilibrium configurations used to pin the spins; then we have to average over the possible different choices of the couplings:
\begin{equation}
\mathcal{F}_E(T,c)=-\frac T N\overline{\sum_{\mathcal{C}_f}P^J({\mathcal{C}_f},T)\log Z^J_{\mathcal{C}_f}(c,T)} \ ,
\end{equation}
where $P^J({\mathcal{C}_f},T)=\exp(-\beta H^J({\mathcal{C}_f}))/Z^J(T)$ is the equilibrium probability of the configuration $\mathcal{C}_f$ in the free system at temperature $T$ ($=1/\beta$); $Z^J_{\mathcal{C}_f}(c,T)$ is the partition function of the system where $cN$ spins at random are constrained to be in the same state than in the reference equilibrium configuration $\mathcal{C}_f$, and $Z^J(T)=\sum_{\mathcal{C}}\exp(-\beta H^J(\mathcal{C}))$ is the partition function for the free system. Note that since the model is fully connected, there is no notion of space and the spin index can always be redefined in such a way that the the blocked spins corresponds the first $cN$ ones. In the following we shall always follow this procedure.  
The average of the free-energy over the configuration $\mathcal{C}_f$ can be simplified introducing $m$ replicas of the system, constrained to have $cN$ identical spins, and sending $m$ to $1$ at the end:
\begin{equation}
\frac{\sum_{\mathcal{C}_f}\exp(-\beta H^J({\mathcal{C}_f}))\log(Z^J_{\mathcal{C}_f})}{\sum_{\mathcal{C}_f}\exp(-\beta H^J(\mathcal{C}_f))} =\lim_{m\rightarrow 1}\frac{1}{m-1}\log\left(\frac{\sum_{\mathcal{C}_f}\exp(-\beta H^J({\mathcal{C}_f}))(Z^J_{\mathcal{C}_f})^{m-1}}{\sum_{\mathcal{C}_f}\exp(-\beta H^J(\mathcal{C}_f))}\right)\ .
\label{mreps}
\end{equation} 
The numerator in the RHS of~(\ref{mreps}) can be rewritten as a partition function, $Z^m_{\Omega_{\mathcal{C}}}$, of $m$ replica constrained to
have the first $cN$ spins identical: 
\begin{equation}
Z^m_{\Omega_{\mathcal{C}}}=\sum_{\mathcal{C}_f}\exp(-\beta H^J({\mathcal{C}_f}))(Z^J_{\mathcal{C}_f})^{m-1}
=\sum_{\mathcal{C}_1,...,\mathcal{C}_m}'\exp\left(-\sum_{i=1}^m\beta H^J({\mathcal{C}_i})\right) \, ,
\end{equation} 
where $\Omega_c$ and the prime above the sum stand for the constraint of having the first $cN$ spins identical for the $m$ replicas.
The average over the quenched disorder is performed introducing, as usual, $n$ replicas and taking the $n\rightarrow0$ limit:
\begin{equation}
\overline{\log Z^m_{\Omega_{\mathcal{C}}}}-\overline{\log Z^J}=\lim_{n\rightarrow 0}\frac{1}{n}[\log \overline{{(Z^m_{\Omega_{\mathcal{C}}})}^n}-\log \overline{(Z^J)^n}] \ .
\label{nreps}
\end{equation} 
Collecting the pieces together, we obtain:
\begin{equation}
\mathcal{F}_E(T,c)=-T\lim_{N\rightarrow\infty}\lim_{n\rightarrow0}\lim_{m\rightarrow1}\frac{\log \overline{{(Z_{\Omega_c}^m)}^n}-\log \overline{(Z^J)^n}}{Nn(m-1)}.
\end{equation}
Standard replica manipulations give the following result:
\begin{equation}
\overline{{(Z_{\Omega_c}^m)}^n}=\int dQ_{a_{\alpha},b_{\beta}}\exp[-NS(Q_{a_{\alpha}b_{\beta}};T,c)]
\label{Zeqpin}
\end{equation}
with $$S(Q_{a_{\alpha}b_{\beta}};c,T)= -\frac{1}{4T^2}\left(\sum_{\alpha}^n\sum_{a_{\alpha}b_{\alpha}}^m(c+(1-c)Q_{a_{\alpha}b_{\alpha}})^p+\sum_{\alpha\neq\beta}^n\sum_{a_{\alpha}b_{\beta}}^mQ_{a_{\alpha}b_{\beta}}^p\right)-\frac{1}{2}(1-c)\log\det(Q_{a_{\alpha}b_{\beta}}) \ ,$$
where the index $a_{\alpha}$ (and $b_{\beta}$) runs over $m$ different values for each of the $n$ possible choices of 
$\alpha$. Replica configurations identified by the same index $\alpha$ have the first $cN$ spins identical, whereas 
replica corresponding to two different values of $\alpha$ have the first $cN$ spins in uncorrelated configurations for 
$T>T_K$. The order parameter $Q_{a_{\alpha}b_{\beta}}$ is the overlap between the $(1-c)N$ free spins of 
replica $a_{\alpha}$ and $b_{\beta}$. The replica structure we focus on, and its physical interpretation, are identical to the one discussed by Monasson\cite{remi}: because of the pinning field induced by the pinned particles the $m$ replicas inside each one of the $n$ groups either fall into the same state, and are characterized by a high value of the 
intra-group overlap or fall into different states and are characterized by a small value of the overlap. Technically, this means that we consider a  replica symmetric {\it ansatz} for the matrix $Q_{a_{\alpha}b_{\beta}}$. We verified 
that taking a more general {\it ansatz}, e.g. 1RSB, is not necessary as long as one focuses on temperatures higher than the glass transition temperature of the unpinned system. This is the regime we study in the following.
Within the RS {\it ansatz}, $Q_{a_{\alpha}b_{\beta}}$ is a block matrix with $n$ diagonal blocks having $m$ diagonal unitary elements and $m$ x $(m-1)$ off-diagonal $q_1$ elements. The off-diagonal blocks have only elements equal to $q_0$.
For this choice of the overlap matrix the action reads as follows:
\begin{eqnarray}
S(q_1,q_0,m;c,T)=&-\frac{\beta^2}{4}n\left[m(m-1)(c+(1-c)q_1)^p-m^2(n-1)q_0^p\right]+
\nonumber\\
&-\frac{1}{2}n(1-c)\left[(m-1)\log(1-q_1)+
\left(1-\frac{1}{n}\right)\log(1-q_1+m(q_1-q_0))+\right.
\nonumber\\
&+\left.\frac{1}{n}\log(1-q_1+m(q_1-q_0)+nmq_0)\right]
\label{blockS}
\end{eqnarray}
\begin{figure}
\centerline{\includegraphics[width=.4\textwidth,angle=-90]{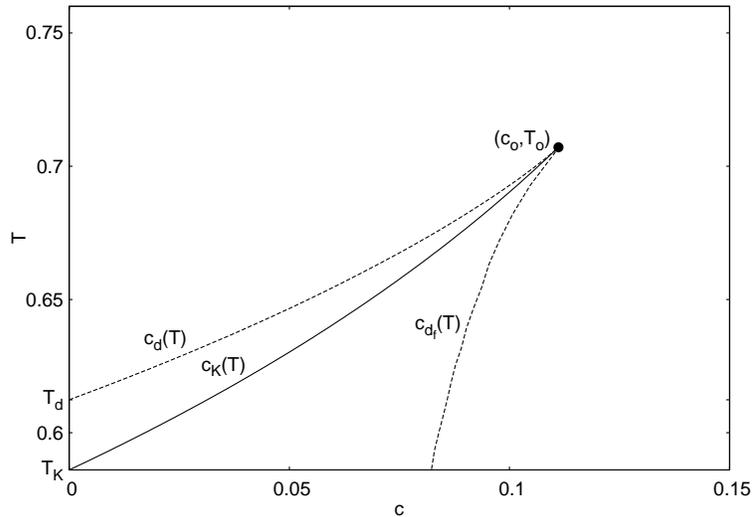}}
\caption{Phase diagram for the spherical $3$-spin model with spins pinned from an equilibrium configuration. The full line represents the thermodynamic transition $c_K(T)$ and the dotted lines correspond to dynamical transitions. The dynamic transition at higher c, $c_d(T)$, is the usual Mode-Coupling transition line found in the study of the equilibrium dynamics. The other one, called $c_{d_f}(T)$, that we report for completeness can be only found through the study of non-equilibrium dynamics, as shown in~\cite{CB_RPGTdyn}. Note that as discussed in the main text, and contrary to the case of completely random pinned spins, it is possible to continuously connect the liquid and the glassy phase through a closed path around the final critical point of the $c_K(T)$ line.}
\label{MFresults_EQ}
\end{figure}
and the saddle point equations over the action in the $n\rightarrow 0$ and $m\rightarrow1$ limit give the following conditions on the parameter $q_1$ and $q_0$:
\begin{eqnarray}
&\frac{p}{2T^2}(c+(1-c)q_1)^{p-1}=\frac{q_1}{1-q_1}
\label{q1}\\
&\frac{p}{2T^2}q_0^{p-1}=\frac{q_0}{(1-q_0)^2}
\label{q0}
\end{eqnarray}
solved by $q_0=0$ and a non-trivial $q_1$. The vanishing of $q_0$ is natural since it measures the overlap between two configurations corresponding to two systems with the same quenched disorder but different ${\mathcal C}_{f}$.  In this case $q_0$ is zero because two typical equilibrium configurations of the unconstrained system have zero overlap above $T_K$ and, hence, the same is expected for the corresponding pinned configurations.

At high temperature or low concentration $c$ of pinned spins this set of equations is solved by a very small $q_1\propto c^{p-1}$.
However, for temperatures lower than $T_d(c)$, a solution characterized by a larger value of $q_1$ appears.
This signals that below this temperature a metastable state lumping together configurations similar to the one used 
to pin particles starts to exist. Thus, the equilibrium dynamical evolution obtained starting from ${\mathcal C}_f$ 
is expected to show an ergodicity breaking. Since ${\mathcal C}_f$, as far as the unpinned spins are concerned, is statistically equivalent to any other equilibrium one, the dynamical behavior starting from any equilibrium configuration for the pinned system is expected to show just the same phenomenon. This allows us to conclude that at $T_d$ 
a dynamical Mode-Coupling-like transition takes place and that 
below $T_d$ the liquid state
of the pinned system is broken in an exponential number of amorphous metastable states.
Indeed, the value of $c_d(T)$ obtained by solving the set of equations constituted by eq. (\ref{q1}) and its derivative
\begin{equation}
\frac{p(p-1)}{2T^2}(1-c)(c+(1-c)q_1)^{p-2}=\frac{1}{(1-q_1)^2} \ ,
\end{equation}
coincides with the one found in the dynamical analysis presented in~\olcite{CB_RPGTdyn}.\\
The secondary minimum, which appears at $c_d(T)$, becomes the global one when:
\begin{equation}
-\frac{1}{2T^2}(c+(1-c)q_1)^p-(1-c)[\log(1-q_1)+q_1]=0 \ .
\end{equation}
This equation defines the line of thermodynamic transition $c_K(T)$. As noticed in the previous paragraph, 
since the metastable state associated to the initial configuration is statistically similar to all other ones, this
means that at $c_K(T)$ the free energy of the liquid state becomes equal to the one of a {\it given} metastable state, {\it
i.e.} the configurational entropy is zero. In consequence, the transition at $c_K$  is
an entropy vanishing transition as the one reached for $T\rightarrow T_K$. For $c>c_K$ the equilibrium phase of the pinned system is given by the state associated to the initial configuration. In this regime, there are still other states but 
they are characterized by a higher free energy. 
In App.~\ref{APP1} we present a more detailed general argument to explain why the equal value of the action at $c_K(T)$ implies  $s_c(c,T)=0$. In App.~\ref{APP3} we directly compute $s_c(c,T)$ by generalizing the Franz and Parisi $\epsilon$-coupling approach to the case of a system with pinned particles and confirm that $s_c(c,T)\downarrow 0$ for $c\uparrow c_K(T)$.\\
The complete phase diagram for the $p=3$-spin model is shown in Fig.~\ref{MFresults_EQ}. 
As shown, the RPGT line stops at a critical point at which the dynamical transitions merge with the static one. 
At this point the glass transition becomes continuous, {\it i.e.} the difference between the high and low overlap values
vanishes.  
The two lines $c_d(T)$ and $c_K(T)$ have been obtained by a numerical solution of the corresponding two sets of equations. They had been already shown in the phase diagram of~\olcite{CB_RPGT} and~\olcite{CB_RPGTdyn}.

\subsection{Random pinning from completely random configurations}

As already stressed before, it is natural to wonder how much the results presented in the previous section depend
on the pinning protocol, in particular what is the difference between pinning from a random versus from an equilibrium configuration. Our expectation, based
on heuristic arguments, was that the difference is substantial \cite{CB_RPGT}. 
In this section we confirm this by repeating the analysis performed above
when $cN$ spins are blocked in a completely random configuration.\\
In this case the average over the reference configuration $\mathcal{C}_f$ is performed over a flat distribution. The computation in this case requires the simple introduction of $n$ replicas all constrained to have the same (random) spin values on $cN$ sites randomly chosen (we used the same notation than in the previous section):
\begin{equation}
\mathcal{F}_R(T,c)=-T\lim_{N\rightarrow\infty}\lim_{n\rightarrow0}\frac{\log \overline{Z_{\Omega_c}^n}}{Nn} \ ,
\end{equation}
where 
\begin{equation}
\overline{Z_{\Omega_c}^n}=\int dQ_{a_{\alpha}b_{\beta}}\exp[-NS(Q_{a_{\alpha}b_{\beta}};T,c)]
\label{Zeqpinr}
\end{equation}
\begin{figure}
\centerline{\includegraphics[width=.4\textwidth,angle=-90]{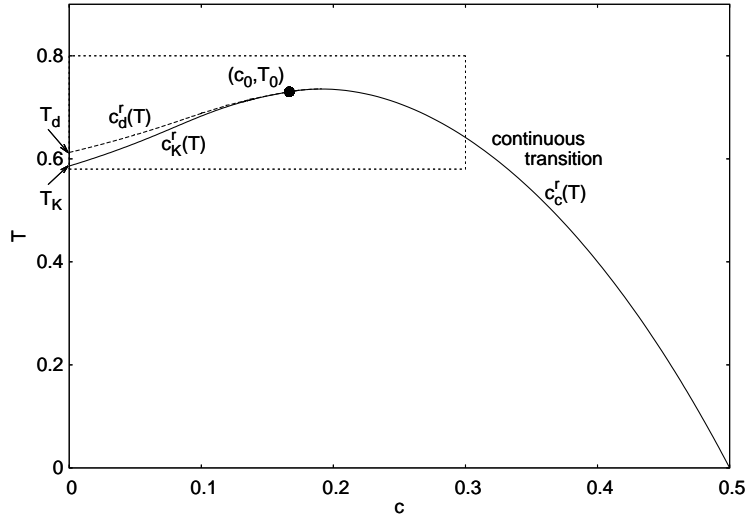}}
\caption{Phase diagram for a spherical $p$-spin ($p=3$) model with spins pinned from a random configuration. The full lines represent the thermodynamic transitions and the dotted line marks the dynamic transition. There is a full low temperature region in the phase diagram characterized by a $1$RSB glass phase. This phase is separated from the liquid phase by a $1$RSB discontinuous-transition line ranging from $(c=0,T_K)$ to a critical point $(c_0,T_0)$ and by a $1$RSB continuous-transition line ranging from this point to zero temperature and $c=0.5$.}
\label{MFresults_RND}
\end{figure}
and
$$S_R(Q_{a_{\alpha}b_{\beta}};c,T)= -\frac{1}{4T^2}\sum_{\alpha\beta}^n\sum_{a_{\alpha}b_{\beta}}^m(c+(1-c)Q_{a_{\alpha}b_{\beta}})^p-\frac{1}{2}(1-c)\log\det(Q_{a_{\alpha}b_{\beta}}) \ .$$
This case of completely random pinned spins is equivalent to a p-spin model in which on each site there is a 
random field that is very strong and equally probable in sign with probability $c$ and that is zero with probability 
$1-c$. As for the p-spin model in a uniform field, it is necessary to consider a 1RSB {\it ansatz}~\cite{CriSom}, where $Q_{a_{\alpha}b_{\beta}}$ presents a block structure with $n/m$ diagonal blocks where the non-unitary values are the off-diagonal $q_1$s and the elements in the off-diagonal blocks are all equal to $q_0$ (we have verified that 
the 1RSB ansatz is stable and there is no need for further RSB). By plugging this specific form of $Q_{a_{\alpha}b_{\beta}}$ into the action we find: 
\begin{eqnarray}
S(q_1,q_0,m;c,T)=&-\frac{\beta^2}{4}n\left[(m-1)(c+(1-c)q_1)^p+m\left(\frac{n}{m}-1\right)(c+(1-c)q_0)^p\right]+
\nonumber\\
&-\frac{1}{2}n(1-c)\left[\left(1-\frac{1}{m}\right)\log(1-q_1)+
\left(\frac{1}{m}-\frac{1}{n}\right)\log(1-q_1+m(q_1-q_0))+\right.
\nonumber\\
&+\left.\frac{1}{n}\log(1-q_1+m(q_1-q_0)+nq_0)\right] \ .
\label{1rsbS}
\end{eqnarray}
The saddle point equations for the parameters $q_1$, $q_0$, and $m$ of the equilibrium matrix are the following:
\begin{eqnarray}
\frac{p}{2T^2}(c+(1-c)q_1)^{p-1}=\frac{1}{m}\left[\frac{1}{1-q_1}-\frac{1}{1-q_1+m(q_1-q_0)}+\frac{mq_0}{[1-q_1+m(q_1-q_0)]^2}\right]\ \ \ \ \ 
\label{q1r}\\
\frac{p}{2T^2}(c+(1-c)q_0)^{p-1}=\frac{q_0}{(1-q_1+m(q_1-q_0))^2}
\label{q0r}\\
\nonumber
\frac{1}{2T^2}[(c+(1-c)q_1)^p-(c+(1-c)q_0)^p]+\frac{(1-c)}{m}\left[\frac{1}{m}\log\left(\frac{1-q_1}{1-q_1+m(q_1-q_0)}\right)+\right. 
\\
\left.+(q_1-q_0)\left(\frac{1}{1-q_1+m(q_1-q_0)}-\frac{mq_0}{(1-q_1+m(q_1-q_0))^2}\right)\right]=0 \ .
\label{mr}
\end{eqnarray}
Since this problem shares many similarities with the problem of a $p$-spin model in a field, it is natural to find, as we do, a non-zero solution for $q_0$ and the presence in the phase diagram, Fig.~\ref{MFresults_RND}, of a continuous transition line between the paramagnetic and the 1RSB phase~\cite{CriSom}.
This line corresponds to the points  in the $c-T$ plane where the previous set of equations admit a solution with $q_1-q_0\rightarrow0$. To obtain the continuous transition line, following the procedure of~\olcite{CriSom}, we developed the full set of equations for small $q_1-q_0$ and we obtained the following conditions on $q_0$, $m$ and $c$ for any $T$:
\begin{eqnarray}
&\frac{p}{2T^2}(c+(1-c)q_0)^{p-1}=\frac{q_0}{(1-q_0)^2}\\
&\frac{p(p-1)}{2T^2}(1-c)(c+(1-c)q_0)^{p-2}=\frac{1}{(1-q_0)^2}\\
&m=(1-q_0)\frac{(p-2)(1-c)}{2(c+(1-c)q_0)} \ .
\end{eqnarray}
By solving these equations in the range where $0\le m\le1$ (corresponding to $(p-2)/(3p-4)\le q_0\le1$) one finds the continuous transition line $c^r_c(T)$.
Out of this range, the transition becomes discontinuous in $q_1-q_0$ and it is always characterized by the value $m=1$.
The equations (\ref{q1r}), (\ref{q0r}), and (\ref{mr}), in the $m\rightarrow1$ limit, define the discontinuous transition line $c^r_K(T)$. In the same limit, also the dynamic transition line $c^r_d(T)$ can be obtained requiring that the derivative of (\ref{q1r}) with respect to $q_1$ is zero instead of satisying equation (\ref{mr}). All results are summarized in the phase diagram of Fig.~\ref{MFresults_RND}. 

\subsection{Discussion on the phase diagrams and comparison between the two pinning procedures}

\begin{figure}
\centerline{\includegraphics[width=0.6\textwidth,angle=0]{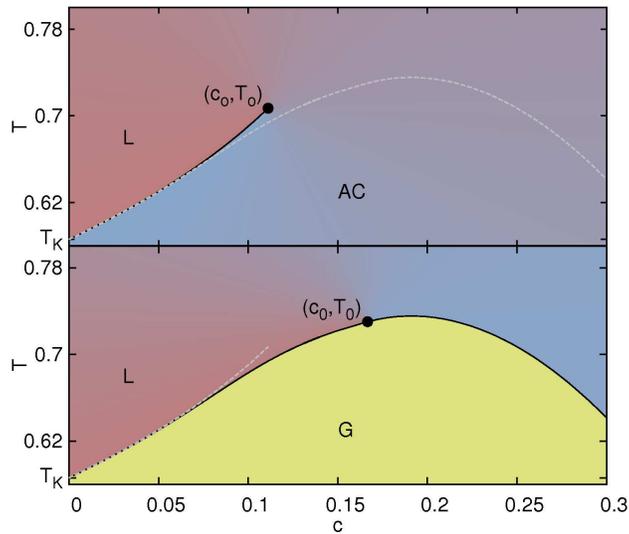}}
\caption{Zoom of the rectangle enclosed by the dashed line in Fig.~\ref{MFresults_RND} and comparison between the thermodynamic glass transitions obtained using the two different protocols to block spins. Top panel: the full line refers to the glass transition induced by pinning spins from an equilibrium configuration; it stops at the critical point $(c_o,T_o)$ at high temperature. Bottom panel: the full line 
refers to the glass transition induced by pinning spins from a completely random configuration. It is a line of $1$-RSB transitions which are discontinuous on the left of the critical point $(c_0,T_0)$ and continuous on the right of it. Below the full line the system is in a one-step replica symmetry broken phase.}
\label{MFcompare}
\end{figure}

We now present a contrastive description of the phase diagrams (see Figs. \ref{MFresults_EQ} and \ref{MFresults_RND}) obtained by the two pinning procedures studied above. Let us first focus on the thermodynamic transition lines. 
In the case of spins pinned from an equilibrium configuration, we obtained a line of thermodynamic glass transitions which separates a liquid phase from an ideal glass phase. These transitions are associated to the vanishing of the configurational entropy in the pinned system.
Two important and related features of these transitions are: (1) the transition line stops in a critical point, $(c_0,T_0)$, where it becomes a continuous transition (2) the glass phase can be obtained from the liquid phase without crossing any thermodynamic singularity.
This is due to the particular kind of low temperature phase of the system, which corresponds to configurations close to the reference amorphous state and not to the usual non-trivial $1$-RSB ideal glass phase. Indeed, the overlap distribution of equilibrium configurations is trivial and characterized by a single peak in the glass phase in contrast to what happens in $1$-RSB phases where it is bimodal, as we shall discuss in more detail later. Even if there is no periodic order in this phase, a non-evident amorphous order is directly imposed by the pinned spins (or particles). As a consequence, to make the difference with usual 1-RSB glasses, we refer to it using    
the evocative name "amorphous crystal" [(AC) in the notation of Fig.~\ref{MFcompare}].
In the case of spins pinned from a random configuration, the glass phase is instead  $1$-RSB and it cannot be reached from the paramagnetic (liquid) phase 
without crossing, either a discontinuous $1$-RSB transition phase, or a continuous one.
This can be understood 
easily since the completely random pinned spins act as an uncorrelated quenched disorder (actually a random 
field) and, hence, the usual RSB glass phases [(G) in the notation of Fig.\ref{MFcompare}], naturally emerges at low temperature\footnote{Actually in this case there is another phase entering in the game; it occupies the right part of the phase diagram, the blue region in the bottom panel. 
We call it frozen liquid. Like the amorphous crystal phase, it is constituted by a stable minimum in the free-energy landscape and can be obtained from the liquid phase without crossing any thermodynamic singularity. On the other hand, it is separated from the glass phase by a continuous transition. Hence metastability and nucleation phenomena does not show up in this case contrary to the other one. Finally it is not characterized by any particular amorphous order but it is by selected by a completely random configuration, {\it i.e.} a liquid configuration at infinite temperature frozen by the pinning particle procedure.}. 
As shown in Fig.~\ref{MFcompare} where we compare the two phase
diagrams, the discontinuous glass transition line emanating from  $c=0,T=T_K$ is quantitatively similar for the two procedures for small values of $c$. Instead, after the critical point $c_0,T_0$ delimiting the final point of the discontinuous transition lines, the behavior is quite different since in the former case the line stops whereas in the latter the line continues until reaching zero temperature for much larger values of $c$.\\
We now focus on the dynamical transitions. In the case of the first pinning procedure one finds 
(see Fig. \ref{MFresults_EQ}) a
Mode Coupling Transition line, $c_d(T)$, and also a new dynamic transition line $c_{d_f}(T)$. The latter 
can only be found in the out of equilibrium dynamics: the relaxation time to equilibrium after quenches from high 
temperature diverges approaching the $c_{d_f}(T)$ line from the right even though the decorrelation time of equilibrium 
correlation functions stays finite, see~\olcite{CB_RPGTdyn}. 
In conclusion, quenches from random configuration lead to aging in the region between the two dynamic transition lines. The situation in the case of spins pinned from a random configuration (Fig. \ref{MFresults_RND} and bottom panel of Fig. \ref{MFcompare}) is quite different. The 
Mode Coupling Transition line, $c_d(T)$, is present also in this case. After the terminating point
$c_0,T_0$, the line coincides with the static continuous one and corresponds to continuous equilibrium dynamical
transitions. Note that the glassy region in which systems quenched from very high temperature remain forever
out of equilibrium is much wider in this case
compared to the previous one.

\section{Renormalization Group analysis of Random Pinning Glass Transitions}
In order to go beyond the mean field analysis and study the critical properties of the glass transitions on the $c_K(T)$ line, we follow the Renormalization Group (RG) procedure proposed in~\olcite{CBTT} and extend it to the case of systems with pinned particles.
This analysis allows one to obtain the divergence of length and time scales that cannot be obtained within  mean field theory. 
\subsection{Field theory description of a replicated liquid}
\label{notationFT}
The starting point of the RG study is a replica field theory based on a Ginzburg-Landau action,
which combines the one found within mean-field theory with a square gradient term that disfavors spatial fluctuations
of the overlap. 
It can be obtained studying the thermodynamics of $m\rightarrow1$ copies (or replicas) of the same liquid system
coupled with a small attractive interaction whose amplitude is set to zero after taking the thermodynamic limit~\cite{remi,MePa_rev}.
It gives a real space description of structural liquids in terms 
of the usual order parameter for the glass transition: the overlap $q_{\mathcal{C},\overline{\mathcal{C}}}$ between an equilibrium configuration $\mathcal{C}$ and a reference configuration $\overline{\mathcal{C}}$, also at equilibrium.
In order to construct a RG theory of structural liquids, one has to consider local fluctuations of the order parameter by introducing a local version of the overlap, $q_{\mathcal{C},\overline{\mathcal{C}}}(r)$ (obtained through a suitable coarse graining over length scales $\Delta$ larger than the inter-particle 
distance, $a_0$\footnote{More precisely it is the microscopic length scale corresponding to the first peak in the radial distribution function.}, 
and smaller than the non-trivial cooperative length scales of the system). 
In the replica language this quantity is represented by the matrix of local overlaps $q_{ab}(r)$ between different replicas $a$ and $b$ of the system.
The simplest Ginzburg-Landau functional obtained retaining the leading terms of the overlap in the field theory and belonging to the universality class of systems showing Random First Order Transition at $T_K$ is the following~\cite{DzSmWo}:
\begin{equation}
\label{eq_GLfunctional}
S\left[ \mathbf q \right] =  \int \frac{d^d x}{a_0^d} \bigg\{\frac{a_0^2}{2} \sum_{a,b=1}^m  \left( \partial q_{ab}(x)\right)^2 +  V(\mathbf{q}(x)) \bigg\}
\end{equation}
where 
\begin{equation}
\label{eq_replica_potential}
V = \sum_{a,b=1}^m \left( \frac{t}{2}q_{ab}^2 -\frac{u+w}{3}q_{ab}^3 + \frac{y}{4}q_{ab}^4 \right)   - \frac{u}{3}\sum_{a,b,c=1}^mq_{ab}q_{bc}q_{ca} \ .
\end{equation}
The temperature dependence in the action is only introduced in the parameter $t=\frac{T-T_0}{T_0}$, where $T_0$ is a suitable scale of the temperature. The other parameters of the potential are positive and independent from the temperature. Compared to the action obtained within mean-field theory, a new space dependent 
term appears in the action to account for the cost of spatial fluctuations of the order parameter.
This real replica description of supercooled liquids allows one to obtain the properties of the metastable states from the knowledge of the replica partition function, 
\begin{equation}
\label{fint}
\mathcal Z(m) = \int \prod_{ab}\mathcal D q_{ab}(x) \exp(-S[\mathbf q]) \ .
\end{equation}
The mean free energy of a typical equilibrium state and the corresponding configurational entropy respectively read~\cite{remi} $\beta F= - \partial \log \mathcal Z(m)/\partial m$ and $s_c=-m^2\partial (\log \mathcal Z(m)/m) /\partial m$. The number $m$ of replicas should be analytically continued to $1$ in the equilibrium liquid phase and to a value less than one in the ideal glass phase~\cite{MePa_rev}.
At the mean-field level, \textit{i.e.} by looking for the uniform saddle-points of (\ref{fint}), one finds that the order parameter $q_{ab}$ is zero above a temperature $T_d$, such that $t_d=\frac{w^2}{4y}$, and that below $T_d$ another uniform solution appears with a  replica symmetric (RS) structure $q_{ab}= q_{EA}>0$ for $a\neq b$. By explicitly using the RS structure of $q_{ab}$ one finds that $q_{EA}$ is the secondary local minimum 
 of $\widetilde V(q)=\left.\frac{V(q)}{m-1}\right|_{m=1}$ and that the configurational entropy per unit volume is $\widetilde V(q_{EA})$.
Finally, at a temperature $T_K$ such that $t_K=\frac{2w^2}{9y}$, there is a random first-order transition with a coexistence between a zero-overlap phase and a high-overlap one, zero latent heat and vanishing configurational entropy density. Below $T_K$,  the system is in an ideal glass phase characterized by a nonzero overlap matrix and a value of $m$ less than $1$.
We have chosen, without much loss of generality, 
a simpler form of $V(q)$~\cite{DzSmWo} which corresponds to impose $q_{EA}=1$ at $T_K$. In this case one finds that
\begin{equation}\label{Vpo}
 \widetilde V(q)=\frac{q^2}{2} \left(\frac{T-\left. T_K\right|_{c=0}}{\left. T_K\right|_{c=0}}+8 \frac{\left. T_d\right|_{c=0}-\left. T_K\right|_{c=0}}{\left. T_K\right|_{c=0}}(1-q)^2 \right) \ . 
 \end{equation}
where by construction $\left. T_d\right|_{c=0}$ and $\left. T_K\right|_{c=0}$ respectively are the mean-field 
dynamical and Kauzmann transition temperatures. The sub-index $c=0$ is introduced for later convenience to recall
that these are characteristic temperatures of the unconstrained super-cooled liquid.\\  
The effect of freezing a fraction $c$ of particles can be schematically included in the field theory by 
forcing the overlap between replica to be equal to $q_{EA}$ at a set of points associated with frozen particles. 
In reality, the effect is more complicated than that but this is irrelevant as far as the large lengthscale properties 
are concerned. Thus, we impose to the measure in (\ref{fint}) the constraints $q_{ab}(x)=q_{EA}$ in 
a random set of Poisson-distributed points characterized by a density $c \rho$, where $\rho$ is the particle density.

\subsection{Migdal-Kadanoff renormalization scheme and mapping to the Random Field Ising Model}
\label{MKRG}
\begin{figure}
\centerline{\includegraphics[width=.49\textwidth]{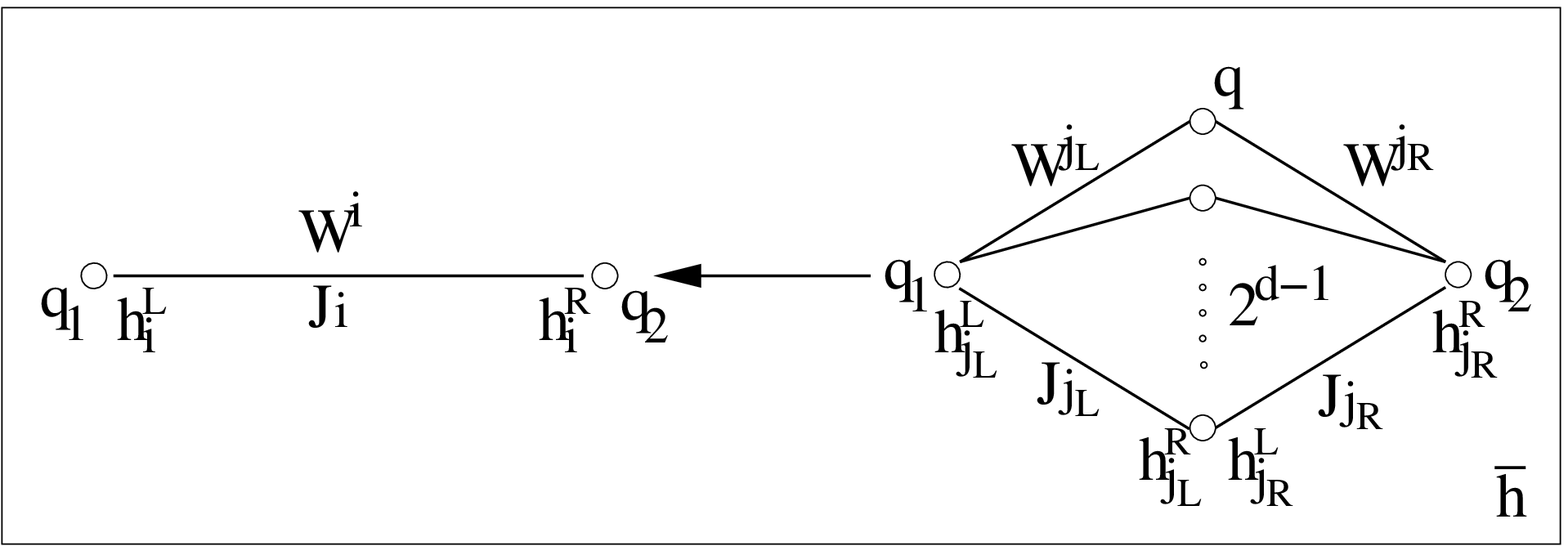}}
\caption{Elementary step illustrating the RG procedure on a hierachical lattice. The notation is explained in the main text.}\label{fig3}
\end{figure}

As discussed in \olcite{CBTT} in order to go beyond the mean-field analysis of RFOT one has to use a non-perturbative
renormalization group approach. Following \olcite{CBTT} we first discretize the field theory
and then apply a Migdal-Kadanoff renormalization scheme. 
This RG method becomes exact on hierarchical diamond-like lattices. Such lattices are built iteratively by replacing each bond between sites by a fixed number of new bonds which, to mimic Euclidean $d$-dimensional lattices, is taken equal to $2^d$. By construction one can then integrate
out iteratively degrees of freedom in the inner bonds and obtain a renormalized effective interaction, see Fig.~\ref{fig3}.
After $n$ iterations, the volume of the system, which is proportional to the total number of original bonds, is renormalized by a factor $2^{nd}$ whereas the ``distance'' on the renormalized lattice corresponds to $2^n$ original bonds: this naturally fixes the length-scale after $n$ iterations as $\ell_n=2^n$.  
In the case without pinned particles at the $n-$th step of renormalization all links are characterized by the
same renormalized effective pair interaction, $W_n^i(\mathbf q^1,\mathbf q^2)$, which satisfies a closed equation written in terms of the pair interactions, $W_{n-1}^{j_L}(\mathbf q^1,\mathbf q^2)$ and $W_{n-1}^{j_R}(\mathbf q^1,\mathbf q^2)$, of the links connecting those sites:
\begin{equation}
\label{eq_integral_recursion}
W_{n}^i(\mathbf q^1 ,\mathbf q^2)=  \sum_{j_{L,R}=1,2^{d-1}}\\\log \int \prod_{a,b}dq_{ab} \exp\bigg\{W_{n-1}^{j_L}(\mathbf q^1,\mathbf q )+  V(\mathbf q)+  W_{n-1}^{j_R}(\mathbf q ,\mathbf q^2)\bigg \}
\end{equation}
where the labels $1$ and $2$ denote the value of two renormalized sites from which emanate $2^{n(d-1)}$ original bonds, and the labels $L$ and $R$ indicate the $j$th left and right link respectively, see Fig.~\ref{fig3}. 
At the $n$th iteration, the original lattice is replaced by a renormalized one where the unit length is $\ell_n$ and 
the pair interaction between sites is $W_n^{i}(\mathbf q^1,\mathbf q^2)$.\\
Freezing particles at random does not change the elementary RG equation (\ref{eq_integral_recursion}) 
but it requires that in the first iterative equations corresponding to $n=1$,  $q_{ab}$ is fixed equal to $q_{EA}$ with probability $c$ for each given intermediate site\footnote{We only fix the constraint for the sites in the center of the hierarchical lattice, corresponding to $n=1$. The reason is that the other sites of the lattice, entering in the RG equations for $n>1$, actually correspond to renormalized regions and not microscopic ones. }. As discussed in the following, the presence of these random constraints considerably complicates 
the analysis because the $W_n$'s become random variables.
In order to obtain a tractable problem we use the insight gained in the work~\olcite{CBTT}.
Without frozen particles we have found that the nature of the RG flow is similar to that found for first-order discontinuity fixed points in which, generically, there are two essential couplings: the field $h$ favoring one 
phase with respect to the other and the coupling $J$ opposing spatial 
variations of the order parameter. 
In our context, $h$ and $J$ respectively correspond to the configurational entropy $s_c$ favoring the zero overlap phase and 
the interface free-energy loss $Y$ between a high-overlap and low-overlap phase. Guided by this result, 
we repeat in the following the analysis performed in~\olcite{CBTT}, "projecting" the functional MKRG equation on simpler ones, where the matrix $q_{ab}$ is a two-state variable, $q_{ab}=0$ or $q_{ab}=q_{EA}$ for all pairs $a,b$.
As in~\olcite{CBTT} we focus on $\widetilde W_{n}^i(q_1,q_2)=\left. \frac{W_{n}^i(q_1,q_2)}{m-1}\right|_{m=1}$ and $\widetilde V(q)=\left. \frac{V(q)}{m-1}\right|_{m=1}$, and we approximate the full integral in (\ref{eq_integral_recursion}) by a steepest-descent calculation (see~\olcite{CBTT} for a discussion of this approximation). The iteration equation simplifies to
\begin{equation}
\label{eq_min_recursion}
\widetilde W_{n}^i(q_1,q_2)=\\ \sum_{j_{L,R}=1,2^{d-1}}\min_{q=0,q_{EA}}\left\lbrace \widetilde W_{n-1}^{j_L}(q_1,q)+\widetilde W_{n-1}^{j_R}(q,q_2) + \widetilde V(q) \right\rbrace , 
\end{equation}
for $q_1$ and $q_2$ that take the values $0$ and $q_{EA}$ only. 
The initial condition is $\widetilde W_{0}^{j_{L,R}}(q_1,q_2)=(q_1-q_2)^2/2$.
We have checked that the results of~\olcite{CBTT} continue to hold within this approximation, as required. 
Because of the frozen particles, the above equation is modified at the first stage of iteration, $n=1$.  
For a fraction $c$ of randomly chosen initial sites\footnote{Note that with a slight abuse of notation we use the same symbol $c$ we used before for the fraction of frozen particles.}, instead of minimizing over $q$, one just evaluates the 
expression within brackets in (\ref{eq_min_recursion}) for $q=q_{EA}$. In consequence, even though the RG equations
do not contain quenched disorder, their initial conditions do. Thus, the $W_{n}^i(q_1,q_2)$s become random variables and
equation (\ref{eq_min_recursion}) defines the flow equation for their probability distributions.    
Actually, it is useful to use the terminology of spin systems and interpret eq. (\ref{eq_min_recursion})
as a minimization equation for a local contribution to the global Hamiltonian. Since a generic pair interaction 
$\widetilde W(q_1,q_2)$ can only take four different values depending on the values of $q_1,q_2$ one can rewrite it, without loss of generality, as 
\[
2\widetilde W(q_1,q_2)=-J_{12}S_1S_2-h_1S_1-h_2S_2+C \qquad S_1=\pm1,S_2=\pm1
\]
where up (down) spins correspond to high (low) local overlaps and $C$ is a constant (the two in the RHS has been
added for consistency with the notation used in \olcite{CBTT}).
Using this dictionary eq. (\ref{eq_min_recursion}) becomes identical to the zero-temperature MK-RG
equations for the random-field Ising model (RFIM) in an external negative magnetic field equal to $-2\widetilde V(q_{EA})$.
The initial condition for the MK-RG equation, $\widetilde W_{0}^{j_{L,R}}(q_1,q_2)=(q_1-q_2)^2/2$, leads to the initial
condition for the flow:
$J_{12}^0=\frac{q^2_{EA}}{2},  h_1^0=0, h_2^0=0$.
At the first iteration, $n=1$, for a fraction $c$ of randomly chosen initial sites, instead of minimizing, 
one just evaluates the expression within brackets in (\ref{eq_min_recursion}) for $S=1$.  This introduces
quenched randomness in the iterative RG equation: it is equivalent to apply a very large 
magnetic field, much larger than the ferromagnetic coupling on randomly chosen sites.
In consequence, we find that the MK-RG equations for the replica field theory maps on the zero-temperature MK-RG 
equation for a RFIM with an unusual random-field, whose distribution is bimodal and asymmetric.
It consists in a peak of weight $1-c$ centered 
on $-2\widetilde V (q_{EA})$ and a peak of weight $c$ centered at a positive value large enough to fix the spins in the up state\footnote{The effect of this contribution it to block a fraction $c$ of spins in the up state, corresponding to the high overlap position. In a finite dimensional lattice one can take for the magnetic field any positive value larger than $J$ times the connectivity of the lattice. In the case of MK-RG any field larger than $2J$ is enough.}. 
In this scheme, the glass transition corresponds to the first-order phase transition between the negative (low overlap) and the positive (high overlap) magnetization state of the RFIM. 
The transition is controlled by the uniform negative field, which is equal to minus the configurational entropy, and the fraction $c$ of frozen up spins.
For each value of the negative magnetic field a transition is induced by changing the number of frozen up spins.
Clearly, the larger is the absolute value of the uniform magnetic field, the larger is the critical value of $c$ needed to counterbalance it.
It is only when the fraction of frozen spins is zero, $c=0$, that the transition takes place at zero magnetic field (i.e. $-2V(q_{EA})=0$).

\begin{figure*}[ht]
\hspace{0.5cm}
\hspace{-2.cm}\subfloat[
1st iteration of RG flow]{\label{fig:1}\includegraphics[width=.4\textwidth,angle=0]{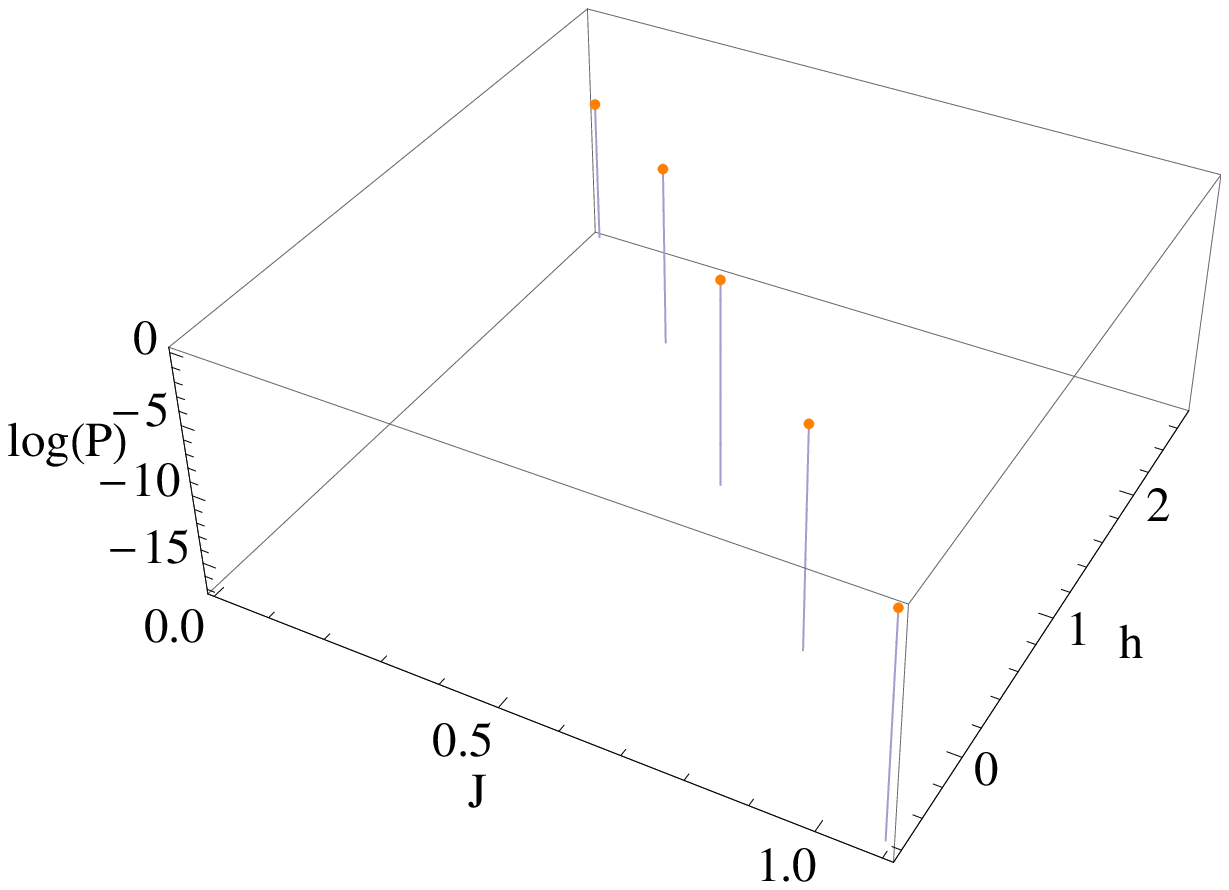}}
\hspace{0.5cm}\subfloat[3rd iteration of RG flow]{\label{fig:2}\includegraphics[width=.42\textwidth,angle=0]{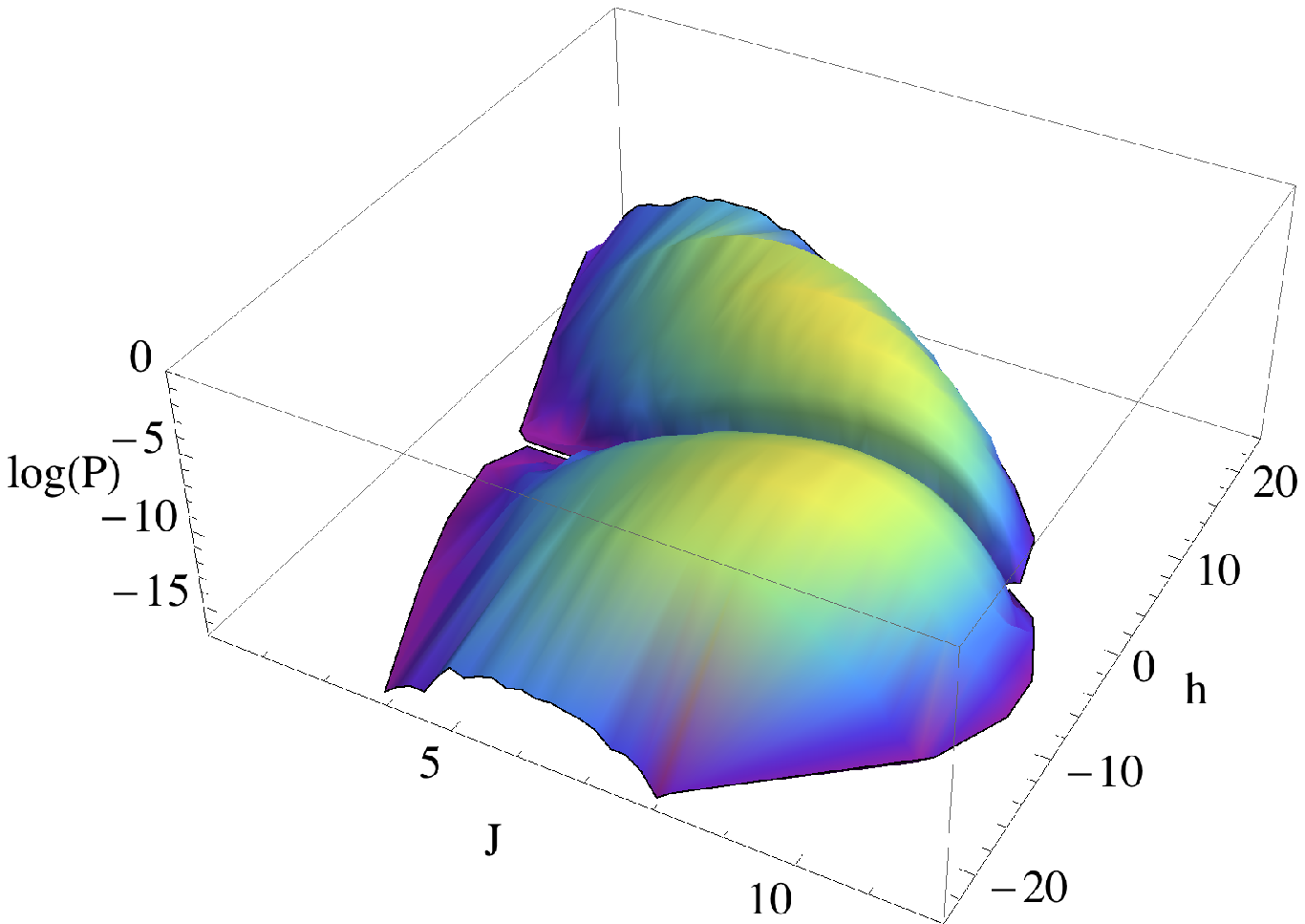}}\\
\hspace{-1.5cm}\subfloat[4th iteration of RG flow]{\label{fig:3}\includegraphics[width=.42\textwidth,angle=0]{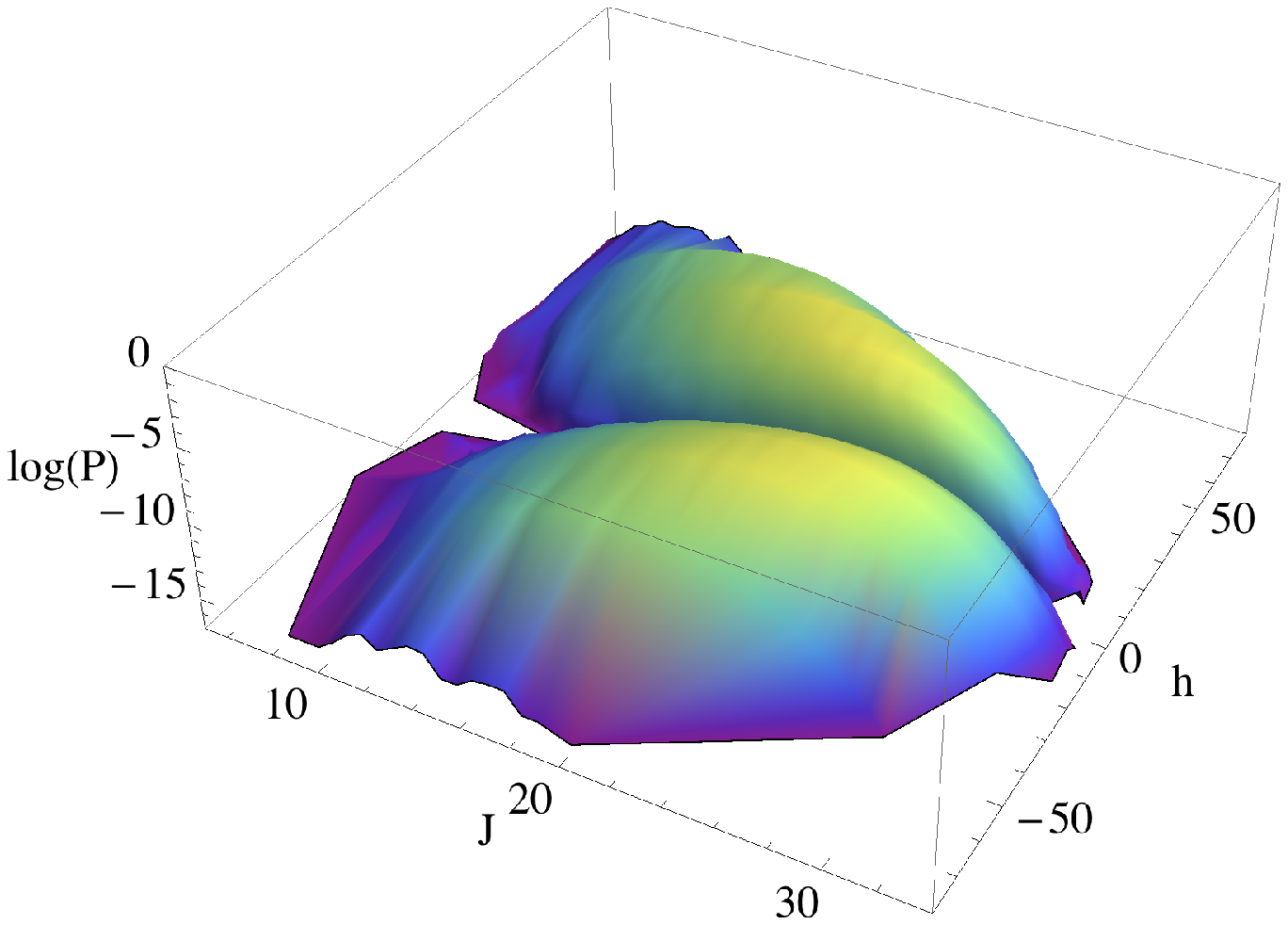}}
\subfloat[13th iteration of RG flow]{\label{fig:4}\includegraphics[width=.46\textwidth,angle=0]{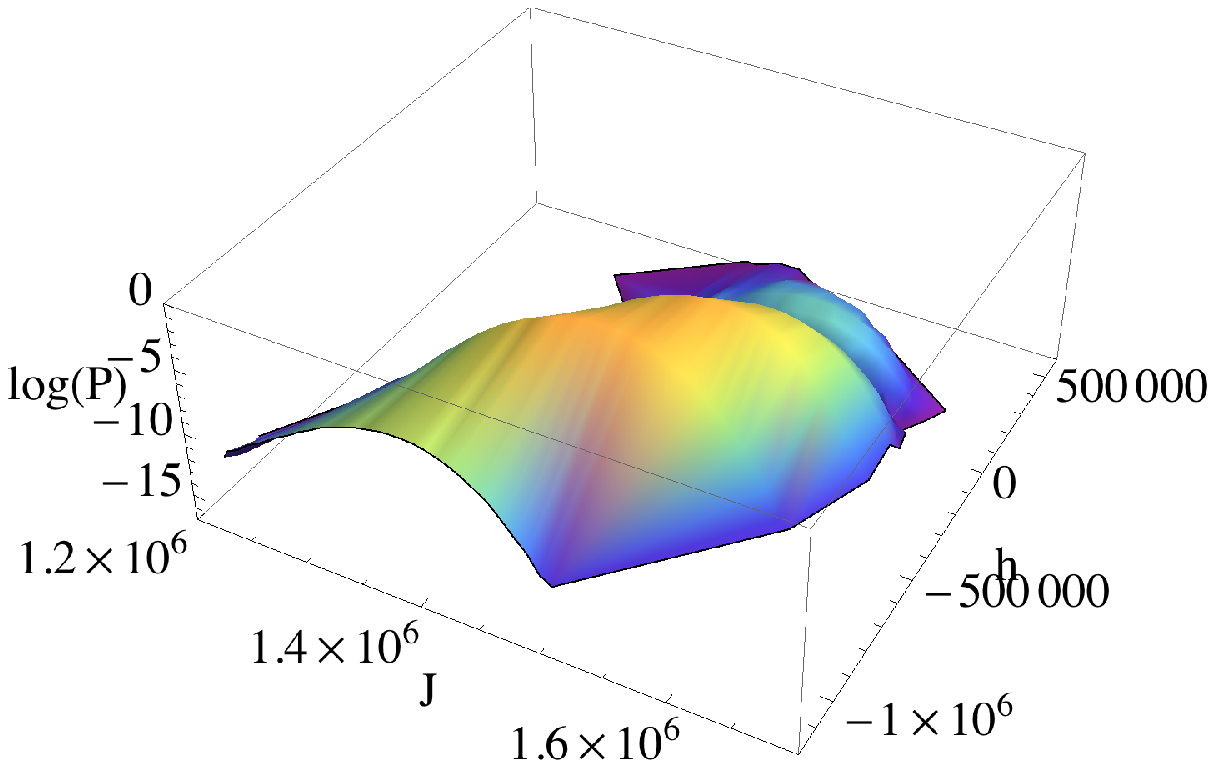}}
\caption{Probability distribution of the coupling constants $J$ and $h$, $\left. P(J,h_L,h_R)\right|_{h_L=h_R=h}$, along the RG flow in the case $\left.T_d\right|_{c=0}/\left.T_K\right|_{c=0}\simeq 1.4$ at $T=1.35\left.T_K\right|_{c=0}$ and $c=0.207435$. The distribution, initially being a pair of delta functions one with weight $1-c$ at small positive field and the other with weight $c$ at a large negative field, is constituted by five delta functions at iteration one (see Panel \ref{fig:1}) where only one delta function corresponds to small positive field (note that its weight is still the largest one) and by a smooth distribution formed by two almost equal parts corresponding to positive and negative fields for iterations larger than three (iteration three in Panel \ref{fig:2}, iteration four in Panel \ref{fig:3}). Eventually one of the two parts shrinks and disappears at $\ell=\ell_{PS}=2^{n^*}$ (iteration thirteen in Panel \ref{fig:4}).}
\label{fig:PDFflow}
\end{figure*}
\subsection{Renormalization Group flow and phase diagram beyond the Mean Field approximation}
\label{discRG}
\begin{figure*}[ht]
\hspace{0.5cm}
\includegraphics[width=.4\textwidth,angle=-90]{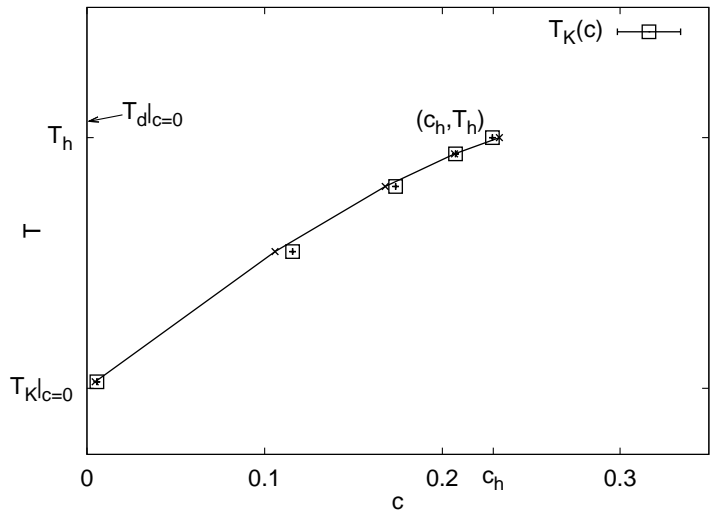}
\caption{Phase diagram obtained by the renormalization group approach for $\frac{\left. T_d\right|_{c=0}}{\left. T_K\right|_{c=0}}=1.4$. The $x$-axis reports the amount of the concentration of pinned particles in the system. The squares are the results obtained by RG for the ideal glass transition line $c_K(T)$, or $T_K(c)$. The continuous line corresponds to the phenomenological result $c_K\propto s_c$. The endpoint of the glass transition line takes place at $c_h,T_h$. Note that 
the temperature $T_h$ is smaller than the mean-field Mode Coupling Transition temperature $\left.T_d\right|_{c=0}$ for the 
unperturbed liquid (found using the Ginzburg Landau action with $\left.T_d\right|_{c=0}/\left.T_K\right|_{c=0}\simeq 1.4$) and, therefore, of the 
value $T_o^{MF}>\left.T_d\right|_{c=0}$ found within mean-field theory.}
\label{3dPhDi}
\end{figure*}

\begin{figure*}[ht]
\includegraphics[width=.7\textwidth,angle=-90]{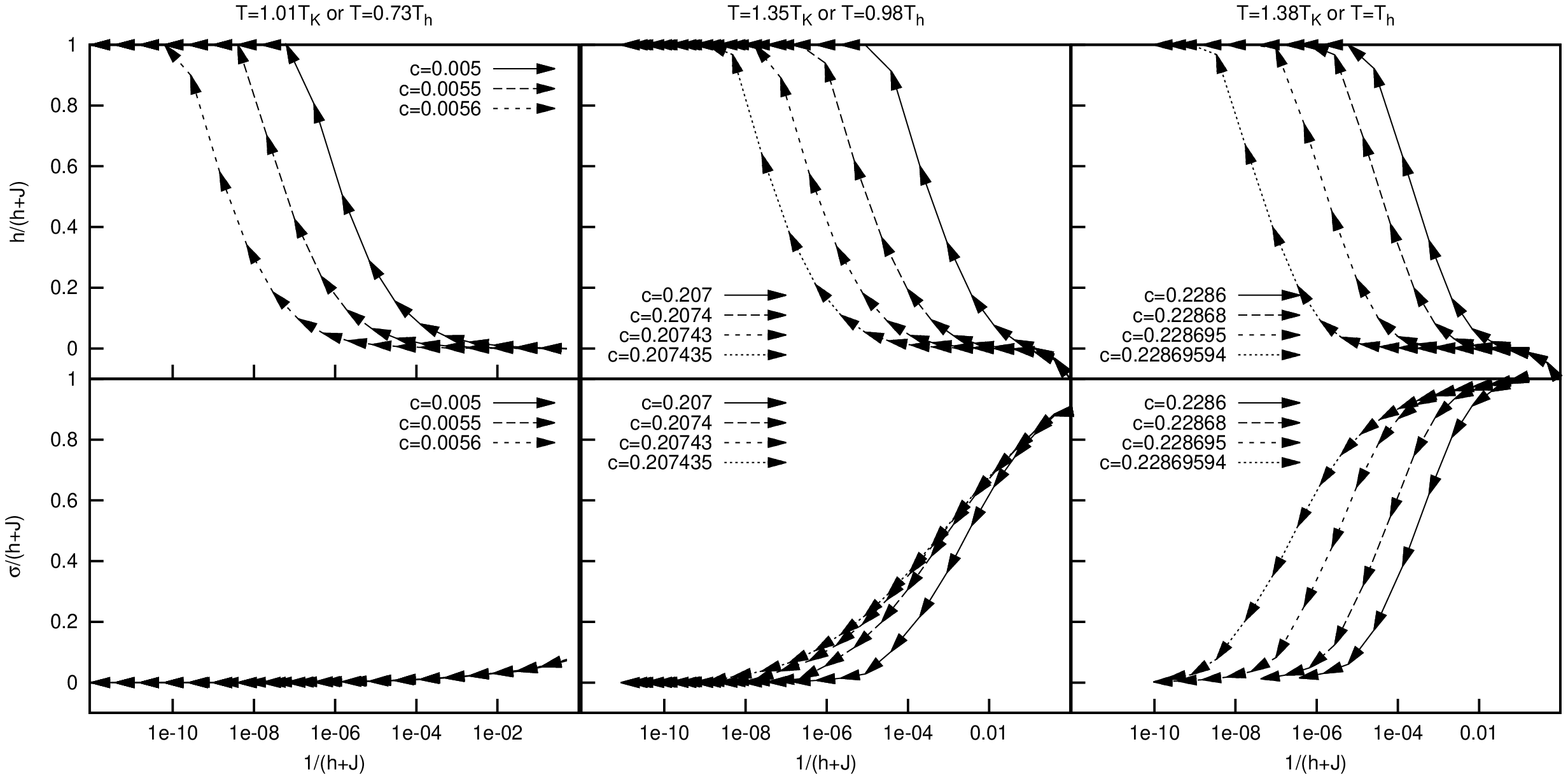}
\vspace{-2.5cm}
\caption{RG flow of the coupling constants approaching the transition, {\it i.e.} for $c_K(T)-c\rightarrow 0^+$, projected on the planes $\sigma=0$ and $h=0$. The left panels correspond to a temperature very close to $\left.T_K\right|_{c=0}$, the central ones to a temperature intermediate between $\left.T_K\right|_{c=0}$ and $T_h$, $T/\left.T_K\right|_{c=0}\simeq1.35$. The right panels correspond to a temperature very close to $T_h$. The reported data, $h,J$, are the average values of $h_n^l,J_n^l$ and $\sigma$ is the variance of $h_n^l$. In the three cases, the projection on the $\sigma=0$ plane leads to the same flows obtained in~\cite{CBTT}: $J$ increases until the point to set length is reached. 
At that point $h$ and $J$ are of the same order. On larger length-scales $J$ decreases rapidly and $h$ continues to increase. In the left panels, the projection of the flow on the $h=0$ plane shows that the value of $\sigma$ remains small: in this regime the disorder does not change the scaling of $J,h$ which are the ones typical of usual first order transition. For the central panels, $\sigma$ always decreases rapidly approaching a decreasing limiting curve for  $c_K(T)-c\rightarrow 0^+$. For $\ell\gg\xi$, when the renormalized $\sigma$ is small enough, the flow recovers the properties due to discontinuity fixed point, see left panels. For the right panels,  the flow is dominated by the RFIM critical point. In this case $\sigma$ first increases as $J$ because the flow is attracted by the zero temperature RFIM fixed point. The limiting curve, for $c_K(T)-c\rightarrow 0^+$ is a constant. This holds until the point to set length, $\ell_{PS}$ is reached. 
}
\label{meavflow}
\end{figure*}

In the literature there are already several analyses of the MK-RG eqs. for the RFIM.
However, the particular random field distribution that results from the mapping discussed in the previous section
was not considered in any of those studies and, hence, required a new analysis that we present in the following. 
We have numerically studied the flow of the joint probability laws of $J$, $h_L$ and $h_R$ by solving the iterative equations by population dynamics \cite{MePaPOP}. 
The initial conditions for the RG flow for a given link are $J=q_{EA}^2/2$ and $h_L=h_R=0$, as discussed in 
the previous section.  At the first step of the RG flow a fraction $c$ of spins are frozen in the up state. This is equivalent to introducing random strong positive local fields that, if present, polarize the central spins in the up position. For any central site, the field is added with a Poisson probability characterized by a first moment equal to $c$. Starting from this random initial condition one then
iterates the MK-RG eqs. We computed renormalized couplings and fields by using populations of $10^7$ variables~\cite{MePaPOP}. \\
Compared to the case studied in \olcite{CBTT} the analysis is complicated by the fact that one has to follow the flow of probability distributions and not just of few coupling constants. 
Fortunately, even though initially these distributions are bimodal with one very sharp peak at the first steps of the RG flow, soon enough they become smooth and broad 
and they can be well characterized by their mean and variance, see  Fig.~\ref{fig:PDFflow}.
The glass transition, like in~\olcite{CBTT}, is found studying when the high overlap--positive magnetization state becomes stable.    
We find that the values of the mean and variance of the couplings and the fields start growing along the renormalization group flow when the temperature is low enough. In the liquid phase there is always a characteristic scale above which replica between different sites decouple because the average coupling eventually becomes smaller than the average field, as shown in Fig.~\ref{meavflow} (top panels) where we plot the ratio $h/(h+J)$.
The iteration $n^*$ at which this happens identifies the characteristic mosaic length-scale $\ell_{PS}=2^{n^*}$, called point-to-set in the glass literature~\cite{CBTT}. The longer replica remain coupled, the closer to the glass transition  the system is and the larger $\ell_{PS}$ is. The line of random pinning glass transitions is obtained studying 
the critical concentration $c_K(T)$ at which $\ell_{PS}$ diverges for any given $T$. In Fig.~\ref{3dPhDi}
we plot the phase diagram obtained for $\frac{\left. T_d\right|_{c=0}}{\left. T_K\right|_{c=0}}=1.4$. We recall that $\left. T_d\right|_{c=0}$ and $\left. T_K\right|_{c=0}$ respectively are the mean-field dynamical and Kauzmann transition temperatures for the unconstrained system obtained from the analysis of $\tilde V(q)$, {\it i.e.} in absence of fluctuations (see eq. \ref{Vpo}). Note that, as discussed in \olcite{CBTT}, $\left. T_K\right|_{c=0}$ is not renormalized by fluctuations because of the saddle-point approximation, whereas $\left. T_d\right|_{c=0}$ does not exist anymore beyond mean-field theory. 
By comparing Fig.~\ref{MFresults_EQ} and Fig.~\ref{3dPhDi} one finds that the mean-field phase diagram is drastically modified by taking into account fluctuations. 
In particular, the endpoint of the RPGTs line, $T_h$, turns out to be smaller than the mean-field mode-coupling transition temperature in absence of pinning, $\left. T_d\right|_{c=0}$, contrary to what happens within mean-field theory, see  Fig.~\ref{MFresults_EQ}. 
Finally, the analysis of liquids with different fragility show that  
the ratio $\frac{T_h}{\left.T_d\right|_{c=0}}$ increases when $\frac{\left.T_d\right|_{c=0}}{\left. T_K\right|_{c=0}}$ decreases, {\it i.e.} for more fragile liquids.\\ 
 The continuous line in Fig.~\ref{3dPhDi} is obtained using the relation $c_K(T)\propto s_c(T)$, which was motivated by 
phenomenological arguments at the beginning of this manuscript. It works surprisingly well. 
The RPGT line ends in a point corresponding to the concentration and the 
temperature $c_h,T_h$. Within the magnetic analogy, the physical reason for this phenomenon
 is that by increasing $T$ the fraction of random spins frozen in the up position, which is needed to counterbalance the large negative uniform magnetic field (the configurational entropy), increases too. The net effect is an increase of the quenched disorder. As it is known, a strong enough disorder destabilizes 
long range order in the RFIM even at zero temperature. The endpoint of the line of RPGTs is precisely related to this effect: above $T_h$ there is no transition between high and low overlap phase because the quenched disorder introduced by random pinning is so strong that long-range order is wiped out. In two dimensions, as it is well known
for the RFIM, even an infinitesimal disorder destroys long-range order. For this reasons no RPGTs are expected in 
dimensions lower than three. 
\\
In the following we discuss thoroughly the critical properties of RPGTs in three dimensional systems. We separate two cases depending on the 
closeness to the endpoint $c_h,T_h$. Far from it the system behavior is controlled by the discontinuity fixed point 
as already found in \olcite{CBTT} without pinning, {\it i.e.} the critical properties at RPGTs are the same
 of usual glass transitions (without pinning). Instead, close to the endpoint the zero temperature RFIM fixed point intervenes 
and controls the critical behavior on time and length-scales that diverges approaching $c_h,T_h$.

\subsubsection{Critical regime I: low pinning}

Let us focus first on temperatures slightly above $T_K$. In this case, by increasing $c$ toward its critical values $c_K(T)$ we find results very similar to the ones obtained decreasing $T$ towards $T_K$ at $c=0$.\\
The average value of the coupling at the $n$-th iteration, $J_n^l$, increases as $Y(c,T) \ell^{d-1}$, where $Y(c,T)$ depend on $T$ and $c$ but remains of order one in the interesting ranges of these two parameters. 
On the other hand, the average value of the field increases as $s_c(c,T)\ell^d$, where $s_c(c,T)$ is by definition the configurational entropy (or the effective magnetic field in spins language) coarse-grained over lengths larger than the distance between pinned degrees of freedom. After some RG iterations the variances of the field and of the coupling become much smaller than the corresponding averages, see lower left panel of Fig.~\ref{meavflow}. This means 
that disorder is irrelevant on large length-scales and, hence, one finds back the same physical behavior 
discussed in \olcite{CBTT} for unconstrained super-cooled liquids.  
In particular, the scalings $Y(c,T) \ell^{d-1}$ and $s_c(c,T)\ell^d$ hold until one reaches values of $\ell$ such that the couplings become of the same order 
of the fields. Above this characteristic length-scale the average field starts to dominate and the average coupling 
decreases (it eventually vanishes for $\ell\rightarrow \infty$). Physically this means that the effective degrees of freedom of the renormalized system become non-interacting. 
This cross-over identifies the point-to-set length scale, $\ell_{PS}$, which physically corresponds to the linear size of regions able to decorrelate from their boundary \cite{CBTT}. Our numerical analysis shows that $s_c(c,T)$ vanishes linearly by increasing $c$ at fixed temperature, {\it i.e.} $s_c(c,T)\propto (c_K(T)-c)$. In consequence the scaling of the point-to-set length is:
\[
\ell_{PS}\propto \frac{1}{c_K(T)-c}
\]
As already discussed in \olcite{CBTT}, on the length-scale $\ell_{PS}$ the renormalized system is like a liquid at its ``onset temperature'' where the (renormalized) point-to-set length is equal to one. The difference with a normal liquid are the values of the renormalized surface tension and configurational entropy, $Y(c)\ell^{d-1}$ and $s_c\ell^d$. These are very large compared to $T$, {\it i.e.} the liquid is at very low $T$ compared to the typical scale of the interaction. 
This suggests that the relaxation time could be obtained by assuming an Arrhenius law
at the scale $\ell_{PS}$ and leads to a generalized Vogel-Fulcher law:  
\[
\log \tau \propto \frac{Y\ell_{PS}^{d-1}}{T} \propto \frac{1}{(c_K(T)-c)^{d-1}}
\]
This critical behavior is valid for all temperatures above $T_K$ and below $T_h$ when 
approaching the RPGT by increasing $c$. However, its region of validity shrinks approaching $T_h$ because 
another fixed point starts to dominate the RG flow. Discussing in detail the increase of time and length scale close
to the endpoint is the aim of the following section. 
 
\subsubsection{Critical regime II: strong pinning and the zero RFIM temperature fixed point}
As previously discussed and shown in the phase diagram, the fraction of pinned particles
needed to induce a RPGT increases with temperature. 
This lead to an enhancement of the quenched disordered and, hence, of the variance of the random field in spins language. Previous works on the RFIM with Gaussian and symmetric bimodal disorder distributions have shown~\cite{MiFi,Aha} that, even at zero temperature, long-range order, and therefore the first-order transition between negative and positive magnetization, are wiped out by a large enough disorder. This is indeed what we find (see the phase diagram): above a certain temperature, $T_h$, the disorder fluctuations induced by freezing particles are strong enough to make the glass transition disappears. The disappearance of long-range order due to the quenched randomness is controlled by a zero temperature fixed point different from the discontinuity one we studied in the previous section. Both are expected to play an important role close to $T_h$, as we shall indeed show in the following. This connection with the RFIM could actually be anticipated on general grounds and has also
been found in a complementary way based on a one-loop RG treatment \cite{FranzParisiRG12}.   

We now present our numerical analysis of the RG flow close to the endpoint of the glass transition line.  
Note that within the mapping to the zero temperature RFIM, the transition at $T_h$ is analogous to the one 
taking place at $\Delta h_c$, the critical value of the variance of the disorder above which no long-range order is possible even at zero temperature.  
As a consequence, as it happens for the RFIM we expect the glass transition to become second order at $T_h$ and $c_K(T_h)$.
Right at this point, the scaling properties are completely dominated by the zero temperature RFIM fixed point. The most important effect
is that the average and the variance of both fields and couplings remain of the same order\cite{FaBeMK}, {\it i.e.} disorder remains relevant on all length-scales. We find that the average values and the variances of $J^l, h^l$ all scale as $\ell^{\theta}$, with $\theta=1.491(4)$ for $d=3$. This is the usual value found by MK-RG for the RFIM~\cite{CaMa}, to be compared to the one obtained by simulations $\theta=1.49(3)$~\cite{MiFi}. 
When $T$ is close, but smaller than $T_h$, the critical properties of the glass transition become quite complicated
since the RG flow is first attracted by the RFIM zero temperature fixed point and 
approaches the standard first-order discontinuity fixed point for very large lengthscales only, see central and 
right panels of Fig 6.

Close to $T_h$ the average value and the variance of the couplings start to grow as $\ell^\theta$. The field instead 
is characterized by a variance of the order $\ell^\theta$ and an average (the configurational entropy) 
that is much smaller but grows faster, as $(c_K(T)-c)\ell^x$ (we use the standard notation for the RFIM exponents \cite{BrayMooreRFIM85}). 
The point to set length is determined as discussed previously as the value of $\ell$ such that the typical renormalized couplings become of the order of the typical fields. This comparison leads to
\[
\ell_{PS}\propto \frac{1}{(c_K(T)-c)^{2/(d+\overline{\eta}-2\eta)}}
\]
where we have used the scaling relation $x-\theta=\frac 1 2 (d+\overline{\eta}-2\eta)$.
By the same argument used in the previous section we also finds a diverging relaxation time-scale:
\[
\log \tau \propto \frac{\ell_{PS}^{\theta}}{T} \propto \frac{1}{(c_K(T)-c)^{2\theta/(d+\overline{\eta}/2-2\eta)}}
\]
Since in d=3 the exponent $\theta$ is very close to $1.5$ and $\overline \eta \simeq 2\eta$ (we indeed
find by MK-RG $x\simeq3$), the relaxation time is expected to grow in this regime almost according to the so called Vogel-Fulcher-Tamman law but with respect to the variable $c_K(T)-c$ instead of $T-T_K$, {\it i.e} $\log \tau \propto (c_K(T)-c)^{-1}$.

This critical behavior described above is valid until $\ell_{PS}$ becomes of the same order of $\xi=(T_h-T)^{-\nu}$ ($\nu$
is the RFIM length-critical exponent), {\it i.e.} if one does not approach $c_K$ close enough. Afterwards, for $\ell_{PS}\gg \xi$ we recover the behavior described in the previous section, which is due to the discontinuity fixed point. In order to understand the interplay between these two regimes 
we use scaling theory and assume that the averages of the coupling and of the field increase respectively  as:
\[
\ell^\theta f_1\left( \frac{\ell}{\xi}\right), \qquad (c_K(T)-c)\ell^xf_2\left( \frac{\ell}{\xi}\right)
\]
This is corroborated by the numerical analysis of the RG flow\footnote{A complete check of this scaling hypothesis is quite difficult numerically since it requires to follow the flow on extremely large length-scales and very close to the endpoint. All evidences we have from the numerics corroborate the scaling assumption. }.
Since the scaling with $\ell$ of the average coupling and of the average field is $\ell^{d-1}$ and $\ell^{d}$
for the discontinuity fixed point, the behavior of the scaling functions has to be $f_1(y)\propto y^{d-1-\theta}$ and 
$f_2(y)\propto y^{d-x}$ for large values of the argument $y$. This implies that eventually, very close to 
$c_K(T)$ for any temperature $T<T_h$, one recovers the scaling discussed in the previous section but with 
singular pre-factors. In particular, the average coupling and the average field increase as $\xi^{\theta-d+1}\ell^{d-1}$ 
and $\xi^{x-d}\ell^d$, leading to divergences of time and length-scales such as:
\[
\ell_{PS}\propto\frac{(T_h-T)^{\nu(x-1-\theta)}}{(c_K(T)-c)},
\qquad
\log \tau \propto \frac{(T_h-T)^\upsilon
}{(c_K(T)-c)^{(d-1)}}
\]
where $\upsilon=\nu(d-\theta-1)+(d-1)(x-1-\theta)\simeq 2.125$ in three dimensions (to obtain this result we used $\nu\simeq2.25$ obtained from previous Migdal Kadanof RG computations\cite{FaBeMK,CaMa}, using instead the numerical result\cite{MiFi} $\nu=1.37$ we have $\upsilon\simeq1.69$). 
Note that close to $T_h$ the MCT relaxation time also diverges with a generalized Vogel-Fulcher law \cite{CB_RPGTdyn}, thus a quite intricated set of cross-overs takes places close to the endpoint of the RPGT line.\\   
The value of the discontinuous jump of the overlap at the transition scales as the magnetization jump for the 
RFIM, i.e. 
\[
\Delta q\propto \xi^{x-d}=(T_h-T)^{\beta}
\]
We recall that because of the scaling relation valid for the RFIM $\beta/\nu=d-x$ \cite{BrayMooreRFIM85}. Since in three dimension\cite{MiFi} $\beta\simeq 0.017$ the transition is expected to still look first-order like even close to $T_h$, except if one is able to reach temperatures extremely close to $T_h$ (and for very large system sizes). 
The cross-over between the two regimes takes place for $\ell_{PS}\simeq \xi$, which corresponds to concentrations of pinned particles such that $(c_K(T)-c)\propto (T_h-T)^{\nu(x-\theta)}$ (where in three dimensions $\nu(x-\theta)\simeq 3.375$  using $\nu=2.25$, or $\nu(x-\theta)\simeq 2$ using $\nu=1.37$).
\section{Mapping to the RFIM and simulations of the three dimensional Ising model with random pinning}
\label{Isingpinning}
We have shown in the previous section that the MK-RG equations for the replica field theory map, after some approximations, to the ones valid at zero temperature for a peculiar random field Ising model characterized by a field
that is negative and uniform but in randomly distributed sites, where is very large and polarizes the spins in the up state.
The RPGT discussed previously corresponds in this setting to the transition from the
negatively magnetized to the positively magnetized state, which becomes stable for a large enough fraction of spins pinned in the up position. When we established the mapping we neglected fluctuations around the saddle point and, as a consequence, we obtained the {\it zero temperature} MK-RG eqs for the RFIM. A heuristic, simplistic but suggestive way of taking into account 
the role of (at least some) fluctuations and to study the dynamical evolution for the pinned system \cite{CB_RPGTdyn} is to consider the same spin model but at finite temperature. Just after pinning, and below 
$c_K(T)$, the liquid has a high overlap with the initial equilibrium configuration; it remains  in this state 
for a long time (diverging for $c\rightarrow c_K(T)$), but it eventually decorrelates from the initial condition and 
equilibrates toward the low overlap state. Similarly, consider an Ising model below $T_c$ and take an equilibrium configuration in presence of
a positive magnetic field $h$; then freeze a fraction $c$ of spins at random and revert the field to $-h$. Now the original configuration is metastable for $c$ smaller than a critical value $c^*$, and stable for larger ones. Despite some important differences, the situation 
is very reminiscent of the one we discussed before for supercooled liquids with $h$ playing the role of the configurational entropy, the positive and negative magnetized states playing the role of 
high and low overlap ones. The life-time of the positively magnetized metastable state, which diverges for $c^*-c \rightarrow 0$,  is analogous, {\it mutatis mutandis}, to the relaxation time that takes to the liquid to decorrelate from the initial configuration and 
that diverges for
$c_K-c \rightarrow 0$. \\
\begin{figure}
\hspace{1.5cm}
\includegraphics[width=0.5\textwidth]{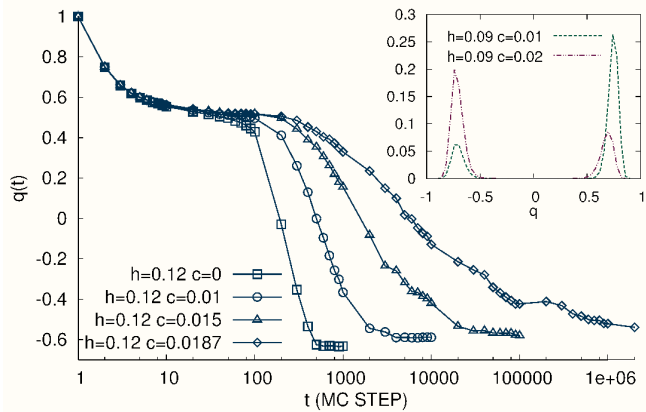}
\caption{Overlap with the reference configuration as a function of time for the 3D Ising model at $T=4$, $h=0.12$ on a cubic lattice with $N=8*10^3$ spins ($J$ is set equal to one). Increasing the fraction of frozen spins a much larger time is needed to escape from
the high overlap phase. Inset: overlap distribution slightly above and below $c^*$  ($T=4$, $h=0.09$, $N=10^3$).}\label{Ct}
\end{figure} 
As benchmark for the super-cooled liquid case, 
we have therefore studied the behavior of the overlap with the initial positively magnetized configuration as a function of time in the corresponding Ising model, see Fig.~\ref{Ct}. 
We indeed found that by increasing the value of $c$ the system spends more and 
more time in the metastable high-overlap state before equilibrating at low overlap. The timescale for relaxation at low overlap diverges
at a critical value $c^*(T,h)$ with the same law discussed before in the glass transition case:  
$\log \tau \propto 1/(c-c^*)^{d-1} $. 
Actually, the curves in Fig.~\ref{Ct} reproduce qualitatively well\footnote{Besides the trivial difference that asymptotic low overlap value is negative.} the ones obtained in super-cooled
liquids by Berthier and Kob~\cite{BeKo_PS}.
This is an indication that the RFOT theory, and the related physical mechanism we have unveiled in~\olcite{CB_RPGT} and in this work,  provide a viable explanation for the mysterious and generic dramatic 
slowing down of the dynamics, which has been observed in simulation when pinning particles of super-cooled liquids \cite{Kim1,Kim2,ProKar11,BeKo_PS}. (The major important difference between this simple Ising model and real liquids is that the original reference configuration is not an equilibrium one. The main consequence is that the system is not automatically at equilibrium after having frozen out particles; thus, relaxation and equilibration time are not the same, contrary to the case of super-cooled liquids.)
In the Ising model case we can also study the equilibrium of the system in proximity of the critical fraction of blocked spins $c^*$. As it is usual for a first order transition, we find a bimodal distribution for the order parameter, which is the overlap between the configuration used to block the spins and any equilibrium configuration, see inset of Fig.~\ref{Ct}. 
The transition point corresponds to perfect bi-modality, whereas instead close to the transition one of the two peaks eventually disappears for large enough system size. This, as we shall discuss in more detail in Sec. 7, suggests a way to study thoroughly RPGTs: exactly at $c_K(T)$ the average probability distribution function of the overlap with the reference configuration used to pin the particles is expected to show a double peak structure of the same kind, while a single peak structure with a peak at low (high) overlap should characterize the probability distribution function at $c<c_K(T)$ ($c>c_K(T)$). 

\section{Comparison with other approaches}
\label{comparison}
In this work and in~\olcite{CB_RPGT} we considered 
one more control parameter besides the temperature in order to favor the glass phase and in order to
give rise to a line of phase transitions having the usual glass transition as one of the two endpoints. 
The very same idea
is also at the root of two other precedent approaches: the ones developed by Franz and Parisi 
\cite{FraPar97} and by Chandler and Garrahan \cite{ChaGar_rev}. Despite their resulting 
phase diagrams look similar at first sight, there are actually very important differences, 
as we explain in the following. 
\subsection{Space-time phase transitions and the $s$-ensemble approach}
Chandler, Garrahan and co-workers proposed to enlarge the number of control parameters 
for a super-cooled liquid by biasing the so-called activity of a given dynamical history~\cite{sensemble,vanwijland,jagach}.
The idea is that the interesting dynamical features in proximity of the glass transition, like the local inhomogeneities in space and time called dynamic heterogeneities, show up in the space-time configurations as a coexistence phenomenon between regions showing high and very low local activity. A Legendre transform of the action with respect to an external variable, $s$, coupled to the activity, is able to provide a quantitative analysis of this phenomenon and to actually reveal the occurrence of a first-order transition in the $s,T$ plane.
The resulting phase diagram is similar to ours in Fig.~\ref{MFresults_EQ} besides the (relatively important) difference that the glass transition of ordinary unbiased super-cooled liquids takes place at zero, and not finite, temperature\footnote{Very recently, it was shown that is also possible that the lower endpoint of the transition line takes place for 
a value of $s$ larger than zero \cite{elmatadjack}.}. Even though the two approaches share this and other similarities, they differ in the 
detailed predictions and the physical content.\\
The most important difference is that the transition on the line in the $s,T$ plane is {\it first-order} (in space-time): 
the dynamical behavior changes discontinuously across the transition without showing any divergence of the
relaxation time. Approaching the glass transition of unbiased super-cooled liquids by decreasing the temperature
is therefore very different from approaching the transition line by increasing $s$.
In our approach, instead, the RPGTs have the same critical properties for $c\uparrow c_K$ of the usual glass transition obtained by decreasing the temperature. In particular the relaxation time diverges approaching the RPGTs by increasing $c$.
Other substantial differences can be found studying limit cases and dimensionality dependence. For instance, 
it has been shown that the space-time phase transition is present also in one dimensional models \cite{ChaGar_rev},
whereas as we pointed out previously the lower critical dimension for the RPGTs is two.\\

\subsection{The $\epsilon$-coupling approach: a detailed comparison}
An approach to the glass transition more similar to the procedure of pinning particle is the one developed by Franz and Parisi~\cite{FraPar97}.
It consists in introducing a bias, $\epsilon$, in the thermodynamics of a glassy system to favor configurations correlated with a reference one, chosen from the equilibrium measure at temperature $T$.
Both the $\epsilon$-coupling and the pinning procedure are equally aimed at reveal 
the complex multi-state structure in the phase space of glassy systems.
In the $\epsilon$-coupling case, as for RPGTs, for every temperature in a finite range $T_K<T<T^*$ it exists is a critical value $\epsilon_s(T)$ for which a thermodynamic transition occurs in the biased system. The resulting 
phase diagram in the  $\epsilon,T$ plane is very much reminiscent of the $c,T$ one (compare the phase diagram in~\olcite{FraPar97} with Fig.~\ref{MFresults_EQ}). 
This notwithstanding, the differences between the two approaches are several.
The most important one is that the transition at $\epsilon_s(T)$ is not a glass transition but is {\it first-order} with non-zero latent-heat \cite{FraPar97} and, hence, is not accompanied by the divergence of the equilibrium correlation time. \\
The physical reason behind these differences is that in the pinning case the metastable state correlated with the 
reference configuration is just one out of the very many in which the system can freeze. It does not differ from them
neither for its internal energy, entropy nor for other properties. Instead,  the $\epsilon$-coupling, introduced by an external term in the Hamiltonian, makes the state correlated with the preferred configuration very special and different from 
all the other ones. For example, the overlap among configurations belonging to the favored minimum is different from the overlap among configurations in any of the other available minima. The internal entropy and even the energy of the favored minimum become extensively different from the ones of other metastable states. 
Thus, whereas by increasing $c$ one reduces the number of possible metastable states until the RPGT where 
the configurational entropy vanishes, this phenomenon is pre-empted in the $\epsilon$-coupling case by 
a first-order transition between the preferred state and all the other possible ones. A recent numerical study has shown that by constraining the overlap to a value intermediate between $q_{EA}$ and zero, the system shows spinodal decomposition, as expected for a first order phase transition \cite{chiaraparisi}. This is an indication that the phase transition predicted with the $\epsilon$-coupling indeed takes place (at least for a finite value of $\epsilon$).

In order to analyze the difference between the two approaches we computed the coupling $\epsilon_K(T)$ that would be able to induce an entropy crisis of the liquid phase. It would represent, in the $\epsilon$-coupling framework, the analogous of the glass transition line $c_K(T)$.
Starting from the results of~\olcite{FraPar97} for the spherical $p(=3)$-spin model and imposing that the low overlap phase must have zero configurational entropy, we obtained the full black line in Fig.~\ref{FrPa_PhDi} (See Appendix C for the computation).
This glass transition line entirely lies in the region where the liquid phase is metastable. In consequence, it is irrelevant as far as equilibrium dynamics (or thermodynamics) is concerned, except for the initial point $(\epsilon_K(T_K)=0,T_K)$ where it coincides with the glass transition in temperature.
In the phase diagram of~\olcite{FraPar97} there are also two more lines very much reminiscent of the dynamic transition lines $c_d(T)$ and $c_{d_f}(T)$ obtained in~\olcite{CB_RPGTdyn} and in this paper, and reported in Fig.~\ref{MFresults_EQ}. 
\begin{figure}
\centerline{\includegraphics[width=.4\textwidth,angle=-90]{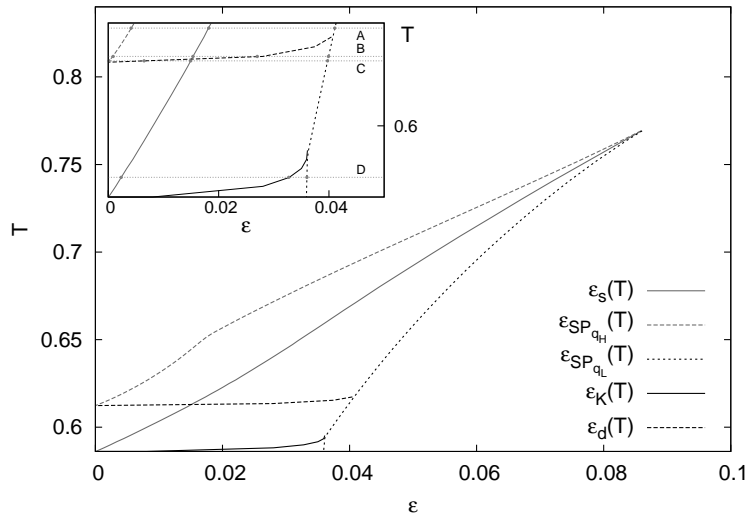}}
\caption{The Franz and Parisi phase diagram~\cite{FraPar97} ($\epsilon_s$, $\epsilon_{SP_{q_H}}$ and $\epsilon_{SP_{q_L}}$ lines) for the $p=3$ spin model has been completed by the introduction of the line of glass transitions for the metastable liquid phase ($\epsilon_K$) and the Mode Coupling transition line ($\epsilon_d$). The inset shows a zoom of the low temperature region of the phase diagram where four ideal experimental paths ($A, B, C$ and $D$) are superimposed to show the possible variety of transitions that it is possible to encounter using the $\epsilon$-coupling set-up.}
\label{FrPa_PhDi}
\end{figure}
However, also in this case appearances are deceiving.
In the $\epsilon$-coupling phase diagram, only the line $\epsilon_{SP_{q_L}}$ reported in Fig.~\ref{FrPa_PhDi}, representing the spinodal points of the liquid phase, is probably the analogous of $c_{d_f}$ below the $\epsilon_d$ line\footnote{This is reasonable if we think that the point where $c_{d_f}$ occurs is the spinodal of amorphous metastable states~\cite{CB_RPGTdyn} and in particular of the threshold metastable states, {\it i.e.} the most numerous metastable states that are placed at relatively high free-energy. In practice, for $c<c_{d_f}$ the non-equilibrium dynamics remains trapped in these threshold states, while for $c>c_{d_f}$ equilibrium in the reference metastable state can be recovered in a finite time. Below the Mode Coupling transition line, $\epsilon_d$, that we will discuss better in the following, the liquid phase is essentially constituted by an assembly of these states, hence spinodal of the liquid phase coincide with the spinodal of the metastable states.
Actually this conclusion is straightforward if metastable states disappear all in once.
Otherwise we could imagine that the spinodal of metastable states at different level of free-energy progressively disappears form high to low free-energy. Even in this case equilibrium will be recovered only when the whole set of metastable states have disappeared hence $c_{d_f}$ conceptually coincides with $\epsilon_{SP_{q_L}}$. However the definition of the line $c_{d_f}$ in~\olcite{CB_RPGTdyn} strictly refers to the disappearance of threshold states. It would be interesting to perform a more detailed analysis to see whether the $c_{d_f}$ line moves when we study non-equilibrium dynamics from initial equilibrium configurations at several intermediate temperature $T'>T$ and not only in the case $T'=\infty$\cite{CB_RPGTdyn}.}. 
At variance, the line $\epsilon_{SP_{q_H}}$ is the line of spinodal points of the reference amorphous metastable state. These points are in general different from the points where generic amorphous metastable states appear in the liquid phase, which correspond to the usual dynamical transition {\it \`a la} Mode Coupling, {\it i.e.} the $c_d(T)$ transition line represented in Fig.~\ref{MFresults_EQ}. We have explicitly computed the MCT line, $\epsilon_d(T)$, and plotted it 
in Fig.~\ref{FrPa_PhDi}. (It was obtained by finding the smallest $\epsilon$ such that equilibrium of the liquid phase is characterized by the presence of a non-trivial solution with $q_1\neq q_0$ and $m=1$, in case of a $1$-RSB {\it ansatz} of the replica matrix.)
The distinction between the two approaches is now evident both in the statics (the transition is first-order and not RFOT like) and in the dynamics (the region where aging is expected, {\it i.e.} the region below $\epsilon_d(T)$, is much smaller in the $\epsilon$-coupling case). \\
The phase diagram for the $\epsilon$-coupling reveals a quite intricate spectrum of transitions when $\epsilon$ increases (see inset of  Fig.~\ref{FrPa_PhDi}).
When $\epsilon$ rises at high temperature (line A), only a first order phase transition  occurs at $\epsilon_s(T)$ between a simple liquid phase and the amorphous state chosen by the reference configuration. This transition is preceded and followed by the spinodal points of the two phases.
For temperatures larger but almost equal to the Mode Coupling temperature of the free system, two different cases are possible: at higher temperature (line B), increasing $\epsilon$, a well defined amorphous reference metastable state forms before the occurrence of a first order transition between the liquid phase and the amorphous state.
However, in this case, a further increasing the coupling between replicas produces in the liquid phase (now metastable) first the formation of a large number of metastable states (the ergodicity of the liquid phase breaks) and finally the disappearance of these new states and of the liquid phase itself.
At lower temperatures still larger than $T_d$ (line C), the first order transition starts to occur after the ergodicity breaking transition of the liquid phase. This implies that eventually a glassy behavior in the equilibrium dynamics becomes visible.
At even lower temperature (line D), finally below $T_d$, the liquid phase is non-ergodic for the free system, yet the thermodynamic transition occurring when $\epsilon$ increases continues to be a first-order transition. Beyond this transition, following the metastable liquid branch, it is now possible to approach at last the true thermodynamic glass transition associated to the vanishing of the configurational entropy.   

\section{On the difference between random pinning procedures and on possible numerical and experimental studies}

The procedure of pinning particles has emerged as a way to probe the complex structure of the phase space of supercooled liquids and to study the occurrence in real systems of ideal glass transitions at $c_K(T)$. 
It can be used to compare and contrast the existing dynamical or thermodynamic theories of glass transition.
In this discussion section we address several general issues related to it.

\subsection{Pinning from an equilibrium configuration versus pinning from a high temperature (random) configuration}

A natural question is how crucial is that the configuration used to pin particles is an equilibrium one.
Actually, pinning from a high temperature (random) configuration is a procedure that has been already studied in the past in connection
with the physics of glass-forming liquids in porous materials, both numerically in simulations \cite{kunipinning} and analytically within MCT \cite{krakoviackpinning}.\\
Our study, based on mean-field glassy systems, show that the two procedures lead to a very different physical
behavior. In the case of RPGTs the phase diagram displays a transition line with an endpoint at a high temperature $T_h$. This implies that it is possible to follow a continuous path in the phase diagram 
connecting the liquid and glass phases without crossing any phase transition, see Fig. \ref{MFresults_EQ}. The main reason that makes this (at first sight) surprisingly fact possible is that the ideal glass phase is "simple": it just consists in the equilibrium configuration from which particles are pinned and vibrations around it. This clarifies why it is not paradoxical 
that liquid and ideal glass are connected by a continuous path and  why the ideal glass phase can be sampled easily.
In the case of pinning from a random configuration the ideal glass phase is instead "non-simple": there is no way to sample it easily and it is separated from the liquid phase by a transition line. This is expected on general grounds since this way of pinning particles (or spins) introduces quenched disorder that is not correlated to any particular state and looks 
completely random (besides possible short-range correlations). Thus, it does not favor any state; instead, it introduces more randomness. The phase diagram, see Fig.~\ref{MFresults_RND}, contains a transition 
line that first, at low $c$, consists in RFOT-like transitions and that is actually similar to the corresponding one of 
RPGTs (except for the differences in the low-temperature phase). However, after the critical point, the transitions change nature and become continuous. This is interesting 
because it predicts that in this case, except at very low pinning, the physical behavior and the criticality are quite different
from the ones displayed or expected for unconstrained super-cooled liquids. In particular there is a very large regime
in which the transition is continuous. Instead of observing the characteristic two-steps behavior for the collective correlation functions one should instead find an overlap $q_1$ that emerges continuously from the long-time limit value $q_0$. 
Moreover, approaching the critical point the spin-glass susceptibility diverges and the relaxation time should increase in a power law way with the distance from the transition, as it happens for spin-glasses. These
are quite remarkable and unusual features for super-cooled liquids. Actually, in a very recent work Karmakar and Parisi \cite{ParKar12} studied the behavior of liquids in presence of particles pinned from a high temperature  (hence, essentially random) configuration and found evidences of a transition different from RFOT and possibly continuous. 
It is certainly worth investigating numerically the properties of this transition and determining 
its universality class. Along the same lines a detailed mean-field study of the general case in which particles are pinned from a configuration equilibrated at temperature $T'$ different from the reference temperature $T$ would be 
useful.

Finally, we would like to mention that the difference we obtained between the physical behaviors induced by the two pinning procedures was in contrast with a prediction obtained by Krakoviack \cite{krakoviackpinning} through an extention of MCT to pinned particles. This was very surprisingly since MCT is thought by many (and by us) to be an essential part of RFOT physics. Thus the predictions we found by mean-field, phenomenological and 
RG arguments were expected to be in agreement with a full fledged MCT computation. In a recent work Flenner and Szamel \cite{szamelpinning} developed 
a different MCT theory for pinned particles in which static and dynamics are treated on equal footing. Their analysis
is in complete agreement with ours, thus solving, at least partially, the difficulty mentioned before.

\subsection{Looking for RPGTs in real systems: physical phenomena and finite size scaling}

An important open issue of the thermodynamic (RFOT) approach to the glass transition is to ascertain the existence of a true transition and determine its properties.
This seemed an hopeless task: because of the dramatic growth of the relaxation
time and the modest increase of the correlation length studying the critical regime is very difficult and, since
the ideal glass phase cannot be reached at equilibrium, showing evidences of an ideal glass transition very challenging. The study of RPGTs provides a new and promising way to tackle this problem and circumvents the difficulties cited above. \\
The main idea is that analyzing and proving the existence of the upper part of the RPGT line is doable (even if numerically challenging) by numerical simulations and, possibly, experiments with optical tweezers. From the knowledge gained in these studies and by extrapolation to lower temperatures 
one can obtain new information on the putative glass transition taking place at $T_K$ for unconstrained super-cooled liquids. 

In the following we discuss peculiar physical phenomena induced by RPGTs and propose ways to
test the existence and the properties of these transitions. By approaching $c_K(T)$ the point-to-set length of the pinned system diverges as a power law, while the relaxation time is conjectured to experience a super-Arrhenius behavior.
Thus, the first basic test is that by increasing $c$ one has to find a really dramatic increase of the 
relaxation time, analogous to the one obtained decreasing the temperature for usual super-cooled liquids.
This has been actually observed in very recent numerical simulations~\cite{BeKo_PS,ProKar11}.
However, this is evidently not enough to assess the existence of a transition at any $c_K(T)$.
\\The great advantage of RPGTs is that the glass phase is "simple": it is characterized by the amorphous order imposed to the system by the configuration used to pin particles. 
Thanks to this property, probing the equilibrium regime beyond the transition is in principle an easy task since 
it is straightforward to start from an equilibrium configuration (the initial configuration used to pin particles). Moreover the equilibrium decorrelation time is not related to collective rearrangements and is given by a microscopic time-scale, which remains finite when $c_K(T)$ is approached from above. At variance, as recalled above, approaching the glass transition from below (using concentrations of pinned particles smaller than $c_K(T)$) gives the usual super-Arrhenius divergence of relaxation time. 
Hence, a clear dynamical signature of the transition, and a way to pinpoint its location in the phase diagram is observing a non-monotonic behavior of the equilibrium decorrelation time when $c$ increases and crosses $c_K(T)$. 
\\Dynamical observations are not enough to certify the existence of RPGTs because, obviously, directly observing the occurrence of a divergence is impossible and the non-monotonic behavior of the relaxation time could be the indication of a simple crossover or an avoided transition. Moreover,  in real particle systems, the microscopic time can be quite large because by pinning particles one 
freezes a lot of vibrations that are instrumental in providing fluctuating free volume that helps single particle motion. 
This effect can mask the presence of a non-monotonous behavior of the relaxation time (a possible solution to avoid this spurious and quite annoying byproduct is, instead of pinning, to restrain the motion of  
a fraction of particles by letting them only vibrate around fixed positions \cite{CGVdyn}.
\\A crucial extra piece of information can be obtained by a systematic study of the thermodynamics and in particular of the overlap $q$ between typical configurations extracted from the equilibrium measure.
Indeed, the presence of a RPGT would imply a sudden change in the distribution of the overlap exactly at $c_K(T)$ from a distribution with a low overlap peak, characteristic of the liquid phase where equilibrium configurations are typically far apart one from the others, to a distribution with a high overlap peak, characteristic of the trivial glass phase correlated with the initial equilibrium configuration. This sudden change is smeared out in systems that can 
be numerically studied in practice because of their finite size. This actually is not necessarily a limitation: a thorough study of finite size effects can provide very valuable information on the critical properties as it happens for standard phase transitions. 
\\We now present our predictions for finite size effect based on the RG study presented in this paper and in~\olcite{CB_RPGT}; we expect three different regimes, which are controlled by the distance from critical point $c_h,T_h$ belonging to the universality class of the RFIM. Let consider the situation in which temperature is close to $T_h$, the case relevant 
for numerical simulations.  Both below and above $T_h$, increasing the system size $L$ for $L<\xi$ (with $\xi$ the RFIM correlation length), the scaling theory of the RFIM implies that the distribution of the overlap is characterized by two peaks whose distance scales as $1/L^{d-x}$. The bi-modality is due to disorder fluctuations and to the simultaneous presence of positively and negatively magnetized regions. Each single peak is broadened by thermal fluctuations on a scale
$1/L^{(d+\eta-2)/2}\ll 1/L^{d-x}$.
Exactly on the transition line (below $T_h$) or on its continuation (above $T_h$) both peaks shrink and increase with $L$ in a similar way.  Instead, if the system is beyond or before the transition line one, of the two peaks get depressed and eventually disappears for large $L$s.
Because of the value of the exponents in three dimensions\cite{MiFi}, $d-x\simeq 0.011$ and $(d+\eta-2)/2\simeq 0.75$, the positions of the peaks decreases very slowly with $L$. In consequence, except if one is able 
to reach very large system sizes, the variation of the peak positions with $L$ should be almost invisible in practice. 
When $L$ becomes of the order of $\xi$ the behavior starts to change depending whether $T>T_h$ or $T<T_h$. In the former case the distribution ceases to be bimodal, hence signaling the absence of the transition. In the latter, instead, the behavior crosses over to the one of long-range ordered systems, {\it i.e.} the distance between the two peaks becomes $L$-independent, of the order of $1/\xi^{d-x}$, and their width decreases as $1/L^{d/2}$. 
\\The concomitance of the dynamic and thermodynamic features discussed above and detailed finite size scaling studies of the probability distribution functions of the overlap provide a new way to ascertain (or disprove) the existence of RPGTs for any $T_K<T<T_h$ and determine its position $c_K(T)$ in the phase diagram.\\
Another interesting and promising research direction that we do not cover here since it goes beyond the scope of this work is studying aging in presence of random pinning. We refer the interested reader to our paper\cite{CB_RPGTdyn} for a mean-field analysis.

\subsection{Dimension and protocol dependence}
Using the insights gained by our RG analysis one can obtain interesting predictions concerning the dimensionality and the protocol dependence. These were already discussed in~\olcite{CB_RPGT}. We report them here for completeness.\\
Because of the mapping on the RFIM, our RG procedure predicts no transition, just a cross-over, in two dimensions.
Thus, the lower critical dimension for RPGTs is two. Before embarking in numerical tests of this prediction it would be interesting to understand the sharpness of this cross-over; we leave this issue for future studies.\\
Of the other pinning procedures introduced in the literature, two-walls, one-wall and the cavity ones, only the first 
can lead to a phase transition akin to the RPGT studied in this work. The reason in the cavity case is clear, the 
system has a finite size, for the one-wall case, see~\olcite{zarinelli}. It is important to realize that in the 
two-wall case the free system has effectively one dimension less, {\it e.g.} in three dimensions the sandwich between
the two walls is two-dimensional (the system is only infinite in the two dimensions parallel to the wall). Because of this,
the lower critical dimension is shifted up of one unity and no transition is expected in three (and lower) dimensions.
This geometry has been studied very recently and, indeed, the corresponding results are compatible with this 
prediction \cite{GrTrCa12}.
\section{Conclusion}
In this work we have presented a detailed study of Random Pinning Glass Transitions that includes mean-field 
and RG analysis. We have obtained the critical properties and discussed possible numerical and experimental tests.
Random pinning appears to provide a new angle to tackle the glass transition problem. We are optimistic that 
in the future, by progressively investigating the physical behavior and the finite size effects for randomly pinned fluids
and studying different ways of pinning particles, a lot of substantial progress will be made. \\\\
\noindent {\bf Acknowledgements}\\
We acknowledge useful discussions with L. Berthier, J.-P. Bouchaud, S. Franz, W. Kob, G. Parisi, G. Tarjus and L. Zdeborova
and support from the ERC grant NPRG-GLASS.\\

\appendix

\section{Entropy vanishing transition for a pinned particle system}
\label{APP1}
One of the advantage of pinning particles from an equilibrium configuration in a supercooled liquid system is the possibility of inducing an entropy vanishing transition at temperature higher than $T_K$.
The existence this kind of transition in real systems and its properties is matter of a longstanding and still open debate.
In this appendix we present a short argument that shows in more details why the transition induced by pinning particles is necessarily associated to the vanishing of configurational entropy.
\\The partition function for a liquid replicated $m$ times and where a fraction $c$ of particles are all constrained to be in the same replica can be schematically written as follows:
\begin{equation}
Z^m=\sum_{\alpha}\exp(-N\beta f_{\alpha})\exp(-N\beta(m-1)f_{res}(f_\alpha))=\exp(N\mathcal{S}(c,T))
\label{Zapp}
\end{equation}
where the sum runs over the metastable states indexed by $\alpha$. 
Only the first replica is allowed to freely explore these metastable states. 
The other $m-1$ replicas are characterized by a reduced free-energy $f_{res}(f)$ and configurational entropy $S_{res}(f)$ due to the degrees of freedom of the remaining free particles.
According to the usual procedure, we can rewrite the replicated partition function as an integral over the levels of free-energy density.
Doing this we have to consider two cases, the first one is the case where the free degrees of freedom of the constrained replicas are still free to explore an exponentially large bunch of metastable states,
in this case we have
\begin{equation}
Z^m=\int df\exp(-N\beta f+Ns_c(f))\exp(-N\beta(m-1)f^{res}(f)+N(m-1)s_c^{res}(f))
\label{Z1app}
\end{equation}
which in the thermodynamic limit, using to the saddle point method, becomes
\begin{equation}
Z^m=\exp(-N\beta f^*+Ns_c(f^*))\exp(-N\beta(m-1)f^{res}(f^*)+N(m-1)s_c^{res}(f^*))
\label{Z1aapp}
\end{equation}
where $f^*$ is such that
\begin{equation}
\beta=\left.\frac{\partial s_c(f)}{\partial f}\right|_{f^*} 
\label{fsapp}
\end{equation}
when $m\rightarrow 1$.
In the second case, the constrained replicas are stuck in the amorphous configuration given by the pinned particles, hence the partition function simply reads
\begin{equation}
Z^m=\int df\exp(-N\beta f+s_c(f))\exp(-N\beta(m-1)f^{res}(f))
\label{Z2app}
\end{equation}
and in the thermodynamics limit it becomes
\begin{equation}
Z^m=\exp(-N\beta f^*+s_c(f^*))\exp(-N\beta(m-1)f^{res}(f^*)) 
\label{Z2aapp}
\end{equation}
where~(\ref{fsapp}) also holds.
The transition between the liquid and the amorphous crystal happens when the total free-energy associated to these two partition functions are equal. This automatically gives at the transition the condition $s_c^{res}(f^*)=0$.

\section{Inducing a glass transition in the $\epsilon$-coupling framework}
\label{APP2}

The $\epsilon,T$ phase diagram~\cite{FraPar97} obtained studying the thermodynamics of a spherical $p$-spin system in presence of an $\epsilon$-coupling between equilibrium configurations apparently shares some features with the $c,T$ phase diagram obtained by pinning particles from an equilibrium configuration.
In this appendix we re-propose the derivations of the two spinodal lines and the thermodynamic transition line of this phase diagram. 
These transitions are actually qualitatively different from the Mode Coupling transition line and the glass transition line obtained by pinning particles from an equilibrium configuration.
Then, for an actual comparison with the $c,T$ phase diagram obtained by pinning particles, we propose an original derivation of the analogous of the Mode Coupling transition line and of the glass transition line in the $\epsilon,T$ phase diagram. 
The novel $\epsilon,T$ phase diagram comes out to be quite different from the $c,T$ one. A detailed comparison between the two final phase diagrams is in the discussion section of this paper.\\
The $\epsilon$-coupling procedure consists in biasing the thermodynamics of a spin system by an external field pointing towards an equilibrium configuration $\mathcal{C}_{ref}$ at temperature $T$. We restrict here the discussion to the case where the temperature of the reference configuration  $\mathcal{C}_{ref}$ is equal to the temperature of the coupled system.
If $\sigma_{ref}$ are the spins in the positions given by the reference configuration, the partition function of the biased system is
\begin{equation}
Z[\epsilon,T,\sigma_{ref}]=\sum_{\{\sigma\}} \exp (-\beta H_J[\sigma]+\beta\epsilon\sigma\sigma_{ref}) \ .
\label{Zapp2}
\end{equation}
Averaging over the possible $\sigma_{ref}$ and the quenched disorder $J$ it is possible to obtain the free-energy of the system
\begin{equation}
\Gamma[\epsilon,T]=-\frac{1}{\beta}\overline{\frac{1}{Z[T]}\sum_{\sigma_{ref}}\exp(-\beta H_J[\sigma_{ref}])\log(Z[\epsilon,T,\sigma_{ref}])}
\label{Gapp2}
\end{equation}
where $$Z[T]=\overline{\sum_{\sigma_{ref}}\exp(-\beta H_J[\sigma_{ref}])} \ ,$$
and its Legendre transform
\begin{equation}
V[q;T]=\min_{\epsilon}\Gamma[\epsilon,T]+\epsilon q-F[T]
\label{Vapp2}
\end{equation}
where $$F[T]=-\frac{1}{\beta}\log(Z[T]) \ .$$
The last function $V[q;T]$ represents the thermodynamic potential associated to every possible value of the similarity $q$ between the reference equilibrium configuration and its coupled configuration. By definition we have imposed $V[0,T]=0$. When a finite $\epsilon$-coupling is imposed, the equilibrium of the system is given by the minimization of $V[q;T]-\epsilon q$, hence by the condition
\begin{equation}
\frac{dV[q;T]}{dq}=\epsilon \ .
\label{STATapp2}
\end{equation}
The averages in~(\ref{Gapp2}) are performed through the replica trick by the introduction of two series of $n$ and $m$ replicas of, respectively, the reference configuration and the coupled system. Suitable overlap matrices among replicas of the two series are consequently used within the following restrictions: a replica symmetric {\it ansatz} for the overlap matrix among replicas of the reference configuration, which means a restriction to the $T>T_K$ region of the phase diagram, and a $1$RSB {\it ansatz} with parameter $q_1,q_0$ and $m\in[0,1]$ for the overlap matrix among replicas of the coupled configurations.
Finally the overlap between a specific reference configuration and every coupled configuration is settled to $q$, while it is assumed to be zero for all the other reference configurations. 
Under these assumptions the potential $V[q;T]$ corresponds to the following expression:
\begin{eqnarray}
\nonumber
V[q;T]=\max_{q_1,q_0,m}-\frac{1}{2\beta}\left\{2\beta^2f(q)-\beta^2[(1-m)f(q_1)+mf(q_0)]+\frac{m-1}{m}\log(1-q_1)\right.\\\left.+\frac{1}{m}\log(1-(1-m)q_1-mq_0)+\frac{q_0-q^2}{1-(1-m)q_1-mq_0}\right\}
\label{Vbisapp2}
\end{eqnarray}
where $f(q)=\frac{1}{2}q^p$, 
and the equations for the stationary points read as follows:
\begin{eqnarray}
\frac{\partial V}{\partial q_0}=0 :& \ q_0-q^2=f'(q_0)[1-(1-m)q_1-mq_0]^2 \label{q0app2}\\
\frac{\partial V}{\partial q_1}=0 :& \ \beta^2(1-m)[f'(q_1)-f'(q_0)]=\frac{(1-m)(q_1-q_0)}{(1-q_1)[1-(1-m)q_1-mq_0]} \label{q1app2}\\
\frac{\partial V}{\partial m}=0 :& \ \beta^2[f(q_1)-f(q_0)]+\frac{1}{m^2}\log\left(\frac{1-q_1}{1-(1-m)q_1-mq_0}\right)+\nonumber\\
& \qquad +\frac{\beta^2}{m}f'(q_1)(1-q_1)-\frac{\beta^2}{m}f'(q_0)[1-(1-m)q_1-mq_0]=0 \label{mapp2}
\end{eqnarray}
where $f'(q)$ is the total derivative of $f(q)$ with respect of $q$.\\
The solution of this set of equations for every value of $q$ gives the parameters $q_1, q_0$ and $m$ to be inserted into (\ref{Vbisapp2}) for finally have the potential $V[q;T]$.
In~\cite{FraPar97} it was found that this potential has a stable minimum in $q=0$ at high temperature and develops at temperature $T_d$ a second minimum at finite overlap $q_{EA}$. 
Further decreasing the temperature, the value of the potential in the secondary minimum decreases and reaches zero for $T=T_s=T_K$. This indicate the occurrence of a thermodynamic transition associated to the vanishing of the configurational entropy. \\
The same features can observed for the system biased by the $\epsilon$-coupling term.
In particular at fixed temperature increasing $\epsilon$, the new potential $V[q;T]-\epsilon q$ has a single minimum at overlap $q_L(\epsilon,T)$ for $0<\epsilon<\epsilon_{SP_{q_H}}(T)$. 
At $\epsilon_{SP_{q_H}}(T)$ a new minimum in correspondence of $q_H(\epsilon,T)$ forms and remains metastable for $\epsilon_{SP_{q_H}}(T)<\epsilon<\epsilon_s(T)$. 
At $\epsilon_s(T)$ a thermodynamic transition occurs, beyond which the equilibrium is given by the high overlap solution between the reference configurations and the configurations coupled to them. Finally in the large coupling regime, for $\epsilon=\epsilon_{SP_{q_L}}(T)$ the minimum at low overlap $q_L(\epsilon,T)$ disappears.
The $\epsilon_s(T)$ line is hence defined by~(\ref{STATapp2}) which has two solutions: $q_L(\epsilon,T)$ and $q_H(\epsilon,T)$, and the condition $\left.(V[q;T]-\epsilon q)\right|_{q_H}=\left.(V[q;T]-\epsilon q)\right|_{q_L}$.
The $\epsilon_{SP_{q_L}}(T)$ and $\epsilon_{SP_{q_H}}(T)$ lines correspond to the $\epsilon$ such that both~(\ref{STATapp2}) and its derivative $d^2V[q;T]/dq^2=0$ are satisfied and for each $T$ this happens for two distinct values of $\epsilon$. \\
In general the transition at $\epsilon_s$ has no reasons to takes place where the configurational entropy vanishes for the coupled system. 
As we will see in the following, the single point where the two transitions coincide is $\epsilon=0,T=T_K$, anywhere else the transition at $\epsilon_s$ is a standard first order transition.
To understand this issue, we have to look at the solution $q_1, q_0$ and $m$ of the set of equations~(\ref{q0app2},\ref{q1app2},\ref{mapp2}) for the stationary point of $V[q;T]$ in correspondence of the stable $q_L(\epsilon_s,T)$. 
These parameters give information on the equilibrium characteristics of the ensemble of configurations coupled by $\epsilon$ to the reference configuration in such a way that the overlap between them and the reference configuration is $q$.  
In general, for every chosen $q$, the parameters $q_1, q_0$ and $m$ change.
For very small $q$, the first two equations of this set have not a solution with $q_1\neq q_0$ for $m=1$, the system coupled to the reference configuration is a simple paramagnet before the Mode Coupling temperature.
When $q$ increases and reaches $q=q^d$, the non-trivial solution $q_1\neq q_0$ for $m=1$ appears 
; this corresponds at the Mode Coupling transition. Beyond this point the phase space of the coupled system is broken up into metastable states with self-overlap $q_1$ and mutual overlap $q_0$.
Finally, the point where the configurational entropy of the coupled system is zero corresponds to $q=q^*$, the point where the solution of the full set of equations gives $m=1$. Beyond this point the full set of equations for $q_1, q_0$ and $m$ gives the result $q_1\neq q_0$ and $m<1$, meaning that the $1$RSB phase has been reached. Hence the point $q^*$ is the extremal point of the $1$RSB region.\\ 
For example in the $3$-spin case, in correspondence of $q_L(\epsilon_s(T),T)$, the system coupled to the reference configuration is very far from the limit $q^*$ of the $1$RSB region and, usually, even from the ergodicity breaking point $q^d$ of the Mode Coupling transition. 
For this reason the transition at $\epsilon_s$ typically represents a first order transition between a trivial paramagnetic phase to the amorphous state pointed by the reference configuration.
To actually find the dynamic and the glass transition line $\epsilon_d(T)$ and $\epsilon_K(T)$ we need to find the $\epsilon$ for which respectively the $q^d$ and the $q^*$ becomes the stationary point $q_L(\epsilon,T)$ of the potential $V[q;T]-\epsilon q$. 
In particular we found a line of $\epsilon_K(T)$ located beyond the line $\epsilon_s(T)$ where $q_L(\epsilon,T)$ becomes metastable, meaning that the coupled paramagnetic phase is never stable when the entropy vanishing transition occurs.
On the contrary, the dynamic transition $\epsilon_d(T)$ is found where the paramagnetic phase is stable only at the lowest temperatures and metastable at higher temperatures.
In summary, in the thermodynamic stable phase of the spherical $p$-spin, the $\epsilon$-coupling procedure is never able to induce a glass transition associated to the vanishing of configurational entropy and only at very low temperatures it is able to induce the dynamic transition due to the ergodicity breaking of phase space of the coupled configurations.

\section{Entropy vanishing transition induced by pinned particles in the $\epsilon$-coupling apporach}
\label{APP3}

The overlap potential $V[q;T]$ obtained in the $\epsilon$-coupling framework provides a valid approach to the glass transition problem. We apply here this description to the case of a $p$-spin model with pinned spins from an equilibrium configuration. We will pay particular attention to the amount of the configurational entropy exactly at the thermodynamic transition $c_K(T)$ to show that this transition is always an entropy vanishing transition.\\
When spins are pinned from an equilibrium configuration the potential between replicas coupled to have overlap $q$ has a structure similar to the one presented in~(\ref{Vbisapp2}) with $f(q)=\frac{1}{2}(c+(1-c)q)^p$: $V[q;c,T]=\max_{q_1,q_0,m}\mathcal{V}[q_1,q_0,m;q;c,T]$ where
\begin{eqnarray}
\nonumber
\mathcal{V}[q_1,q_0,m;q;c,T]=\\
\nonumber
\hspace{0.5cm}-\frac{1}{2\beta}\left\{2\beta^2f(q)-\beta^2[(1-m)f(q_1)+mf(q_0)]+(1-c)\frac{m-1}{m}\log(1-q_1)\right.\\
\hspace{1.1cm}\left.+(1-c)\frac{1}{m}\log(1-(1-m)q_1-mq_0)+(1-c)\frac{q_0-q^2}{1-(1-m)q_1-mq_0}\right\} \ .
\label{action}
\end{eqnarray}
The points $c_K(T)$ where the static transition occurs satisfy 
the condition $V[q_L;c,T]=V[q_H;c,T]$, where $q_L(c,T)$ and $q_H(c,T)$ are such that $\partial V/\partial q|_{q_L}=0$ and $\partial V/\partial q|_{q_H}=0$.
Note that, first, $V[q;c,T]$ is the result of a maximization of the action with respect to $q_0$, $q_1$ and $m$. The equations for finding the stationary point in $q_1$, $q_0$ and $m$ have to be satisfied by the $m=1$ condition when the glass transition occurs.
Second, this condition is equivalent to impose that the configurational entropy $s_c=\partial \mathcal{V}/\partial m|_{m=1, q=q_L}$ of the coupled configurations vanishes at the transition.
We now show that the point where $s_c$ vanishes coincide with the point $V[q_L;c,T]=V[q_H;c,T]$ where the static transition occurs.\\
The stationary point conditions over $q_1$, $q_0$ and $m$ with the condition $m=1$ determine the glass transition point and read as follows:
\begin{eqnarray}
\label{pinstationary1}
\frac{q_1-q_0}{(1-q_1)(1-q_0)}=\beta^2\frac{3}{2}\left[[c+(1-c)q_1]^2-[c+(1-c)q_0]^2\right]\\
\frac{q_0-q^2}{(1-q_0)^2}=\beta^2\frac{3}{2}[c+(1-c)q_0]^2
\label{pinstationary2} \\
\nonumber
s_c=\frac{\beta^2}{2}\left[[c+(1-c)q_1]^3-[c+(1-c)q_0]^3\right]+(1-c)\log\left(\frac{1-q_1}{1-q_0}\right)+\\
\hspace{5.cm}+(1-c)\frac{(q_1-q_0)(1-2q_0+q^2)}{(1-q_0)^2}=0 \ .
\label{pinstationary3} 
\end{eqnarray}
The solution $q_1^{L,H}$, $q_0^{L,H}$ to this set of equations has to be found in correspondence of the stationary points $q_{L,H}(c,T)$ that satisfy the following condition
\begin{equation}
\beta^2\frac{3}{2}[c+(1-c)q_{L,H}]^2=\frac{q_{L,H}}{1-q^{L,H}_0}
\label{pinqstationary}
\end{equation}
obtained by imposing $\partial \mathcal{V}/\partial q|_{q_{L,H}}=0$.\\
We must show that the glass transition point corresponds to the point $\mathcal{V}[q_1^L,q_0^L,m=1;q_L;c,T]=\mathcal{V}[q_1^H,q_0^H,m=1;q_H;c,T]$ where the potential shows a thermodynamic transition with a discontinuous jump in the overlap $q$ from $q_L$ to $q_H$.\\
Two tentative solutions for the set of equations~(\ref{pinstationary1}), (\ref{pinstationary2}) and (\ref{pinqstationary}) are 
$q_H=q_0^H=q_1^H$ and $q_L=q_0^L\neq q_1^L$. Indeed, when $q_0^{L,H}=q_{L,H}$, the conditions~(\ref{pinstationary2}) and~(\ref{pinqstationary}) correspond to each other. Moreover the relation~(\ref{pinstationary1}) is identically satisfied for the solution $q=q_H$ while it remains as a constraint to $q_1^L=q_1(q_L,c,T)$ for the solution $q=q_L$:
\begin{equation}
\frac{q_1^L-q_L}{(1-q_1^L)(1-q_L)}=\beta^2\frac{3}{2}\left[[c+(1-c)q_1^L]^2-[c+(1-c)q_L]^2\right] \ .
\label{pinstationary1bis}
\end{equation}
Finally, the condition for the static transition $\mathcal{V}[q_1^L,q_0^L,m=1;q_L;c,T]=\mathcal{V}[q_1^H,q_0^H,m=1;q_H;c,T]$ is 
\begin{eqnarray}
\nonumber
-\frac{1}{2\beta}\left[\frac{\beta^2}{2}[c+(1-c)q_H]^3+(1-c)\log(1-q_H)+(1-c)q_H\right]=\\
\label{VqHVqL}
-\frac{1}{2\beta}\left[\frac{\beta^2}{2}[c+(1-c)q_L]^3+(1-c)\log(1-q_L)+(1-c)q_L\right] \ .
\end{eqnarray}
On the other hand, for the low solution, $q=q_L$, the configurational entropy reads 
\begin{eqnarray}
\nonumber
s_c=\frac{\beta^2}{2}\left[[c+(1-c)q_1^L]^3-[c+(1-c)q_L]^3\right]+(1-c)\log\left(\frac{1-q_1^L}{1-q_L}\right)+\\
\hspace{8.cm}+(1-c)(q_1^L-q_L) \ .
\label{pinstationary3m1qL}
\end{eqnarray}
Comparing the configurational entropy expression with~(\ref{VqHVqL}), it is evident that we would have $s_c=0$ at the static transition, if $q_1^L=q_H$.
Indeed, this is exactly what happens: coming back to~(\ref{pinstationary1bis}) and using~(\ref{pinqstationary}), we obtain that $q_1^L$ obeys to the same equation~(\ref{pinqstationary}) as $q_H$.
Hence we showed that the static transition is always accompanied by the vanishing of the configurational entropy.\\
This particular symmetry in the stationary conditions for the potential $\mathcal{V}[q_1,q_0,m;q;c,T]$ is obviously also present in the free $p$-spin model that can be obtained from the analyzed case putting $c=0$. This is at the origin of the usual glass transition at $T_K$.
At variance, this symmetry gets immediately broken when an external $\epsilon$ coupling between replicas is added to the Hamiltonian. In this case, the previous set of values for the overlap $q_H=q_0^H=q_1^H=q_1^L\neq q_L=q_0^L$ does not constitute a solution for the problem. Instead slightly different values of $q$s characterize $q_H$, $q_0^H$ and the high overlaps $q_1$ in correspondence of $q_L$ and $q_H$. Moreover an explicit term $-\epsilon q$ makes the condition $\mathcal{V}[q_1^L,q_0^L,m=1;q_L;c,T]=\mathcal{V}[q_1^H,q_0^H,m=1;q_H;c,T]$ unmistakably different from~(\ref{pinstationary3}) where $\epsilon$ does not appear. This indicates that the static transition induced by the introduction of any finite coupling between replicas is different both from the glass transition at $T_K$ and the glass transition induced at higher values of the temperature by pinning particles.


\end{document}